\documentclass[sigconf,nonacm]{acmart}
\usepackage{lipsum}
\usepackage{subcaption}
\usepackage{macros}
\usepackage{csquotes}
\usepackage{makecell}
\usepackage{fancybox}
\usepackage{multirow}
\usepackage{colortbl}
\usepackage{array}
\usepackage{xspace}
\usepackage{tikz}
\usepackage{algorithm}
\usepackage{algorithmic}
\usepackage{enumitem}

\usepackage{graphicx}

\usepackage{makecell}

\usepackage{booktabs, xcolor, ragged2e}
\definecolor{ktblue}{RGB}{26,95,168}
\definecolor{gptbrown}{RGB}{184,92,0}
\newcolumntype{P}[1]{>{\raggedright\arraybackslash}p{#1}}

\let\oldfbox\fbox
\renewcommand{\fbox}[1]{{\color{gray}\oldfbox{\color{black}#1}}}

\newcommand{\chip}[2]{%
  \tikz[baseline=(text.base), scale=0.75]{%
    \node[
      draw=black!20,
      fill=#1!5,
      rounded corners=1.5pt,
      inner xsep=1pt,
      inner ysep=0.15pt,
      text=black
    ] (text) {\strut #2};%
  }%
}

\newcommand{\node}[2]{%
  \tikz[baseline=(text.base), scale=0.75]{%
    \node[
      draw=#1!70,
      fill=#1!70,
      rounded corners=1.5pt,
      inner xsep=1pt,
      inner ysep=0.15pt,
      text=black
    ] (text) {\strut #2};%
  }%
}

\newcommand{\pill}[1]{%
  \tikz[baseline=-0.6ex]{%
    \node[
      fill=#1!90,
      rounded corners=2pt,
      minimum width=0.6em,
      minimum height=0.6em
    ] (text) {};%
  }%
}

\definecolor{ButtonColor}{HTML}{57534e} 
\definecolor{DomainColor}{HTML}{fcf8e8} 
\definecolor{PaperColor}{HTML}{57534e} 
\definecolor{PapersColor}{HTML}{15803d}
\definecolor{ReportColor}{HTML}{7e22ce} 
\definecolor{PlanColor}{HTML}{1d4ed8}

\definecolor{LegendTheory}{HTML}{FF8181}
\definecolor{LegendDiscovery}{HTML}{A4CCFF}
\definecolor{LegendInvention}{HTML}{FFC67A}
\definecolor{LegendPolitics}{HTML}{64E06B}
\definecolor{LegendArt}{HTML}{FF94DB}
\definecolor{LegendEconomics}{HTML}{A4CCFF}
\definecolor{LegendOther}{HTML}{EAEAEA}
\definecolor{TabColor}{HTML}{0393D0}

\definecolor{events}{HTML}{f1bfb3}
\definecolor{explain}{HTML}{c8d0c8}
\definecolor{relationships}{HTML}{b2cee6}
\definecolor{button}{HTML}{d9dadb}
\definecolor{LIGHTGRAY}{HTML}{DDDDDD}
\definecolor{REDORANGE}{HTML}{FF4500}
\definecolor{YELLOWGREEN}{HTML}{9ACD32}
\definecolor{PURPLE}{HTML}{9ACD32}
\definecolor{RED}{rgb}{0.7,0.0,0.0}

\definecolor{TeleportationPattern}{HTML}{944BF7}
\definecolor{ProgressivePattern}{HTML}{54B9AF}
\definecolor{OscillationPattern}{HTML}{D9885D}
\definecolor{ProgressivePattern}{HTML}{54B9AF}
\definecolor{ConvergencePattern}{HTML}{FB6B6D}
\definecolor{MultiEraPattern}{HTML}{6289EE}

\newcommand{\events}{\chip{ButtonColor}{\textsf{Events}}\xspace}
\newcommand{\explain}{\chip{ButtonColor}{\textsf{Explain}}\xspace}
\newcommand{\questions}{\chip{ButtonColor}{\textsf{Questions}}\xspace}
\newcommand{\relationships}{\chip{ButtonColor}{\textsf{Generate Relationships}}\xspace}

\newcommand{\tab}[2]{%
  \tikz[baseline=(text.base), scale=0.75]{%
    \node[
      draw=black!20,
      fill=white,
      rounded corners=1.5pt,
      inner xsep=1pt,
      inner ysep=0.15pt,
      text=#1
    ] (text) {\strut #2};%
  }%
}

\newcommand{\hlcb}[2]{%
  \begingroup
  \setlength{\fboxsep}{0pt}
  \raisebox{0pt}[0.9ex][0.5ex]{\fcolorbox{#1!80!black}{#1!20}{\strut #2}}%
  \endgroup
}

\AtBeginDocument{%
  \providecommand\BibTeX{{%
    \normalfont B\kern-0.5em{\scshape i\kern-0.25em b}\kern-0.8em\TeX}}}

\begin{document}






\title[KnowledgeTrail]{KnowledgeTrail: Understanding Patterns of \\Exploratory Sensemaking and Serendipitous Discovery \\in Generative Timelines}

\newcommand{\sys}[0]{KnowledgeTrail}

\author{Sangho Suh}
\orcid{0000-0003-4617-5116}
\affiliation{%
  \institution{Allen Institute for AI}
  \city{Seattle}
  \country{USA}
}
\email{sanghos@allenai.org}

\author{Rahul Hingorani}
\affiliation{%
  \institution{University of Toronto}
  \city{Toronto}
  \country{Canada}
}
\email{rahul.hingorani@mail.utoronto.ca}

\author{Bryan Wang}
\orcid{0000-0001-9016-038X}
\affiliation{%
  \institution{Adobe Research}
  \city{Seattle}
  \country{USA}
}
\email{bryanw@adobe.com}

\author{Tovi Grossman}
\affiliation{
  \institution{University of Toronto}
  \city{Toronto}
  \country{Canada}
}
\email{tovi@dgp.toronto.edu}

\begin{abstract}
Generative AI is reshaping how interactive systems support the discovery and understanding of knowledge, such as how events and ideas developed over time. Yet timelines, one of the most fundamental tools for enabling this kind of temporal sensemaking, remain largely static, limiting user agency and curiosity-driven exploration. Moreover, little is known about how people may explore and make sense of information if the structure of a timeline itself can evolve through interaction. To investigate this question, we designed KnowledgeTrail, a generative timeline that grows and adapts to users' questions, expanding and contracting with their evolving curiosity. Through an exploratory study, we examined how people engage with KnowledgeTrail and identified several patterns of curiosity-driven exploration, serendipitous discovery, and relational sensemaking that differ from conversational AI interfaces. Based on these analyses, our work contributes design implications for future generative systems that support dynamic, user-driven tracing of ideas, events, and their relationships.
\end{abstract}

\begin{teaserfigure}
    \centering
    \includegraphics[trim=0cm 0cm 0cm 0cm, clip=true, width=\textwidth]{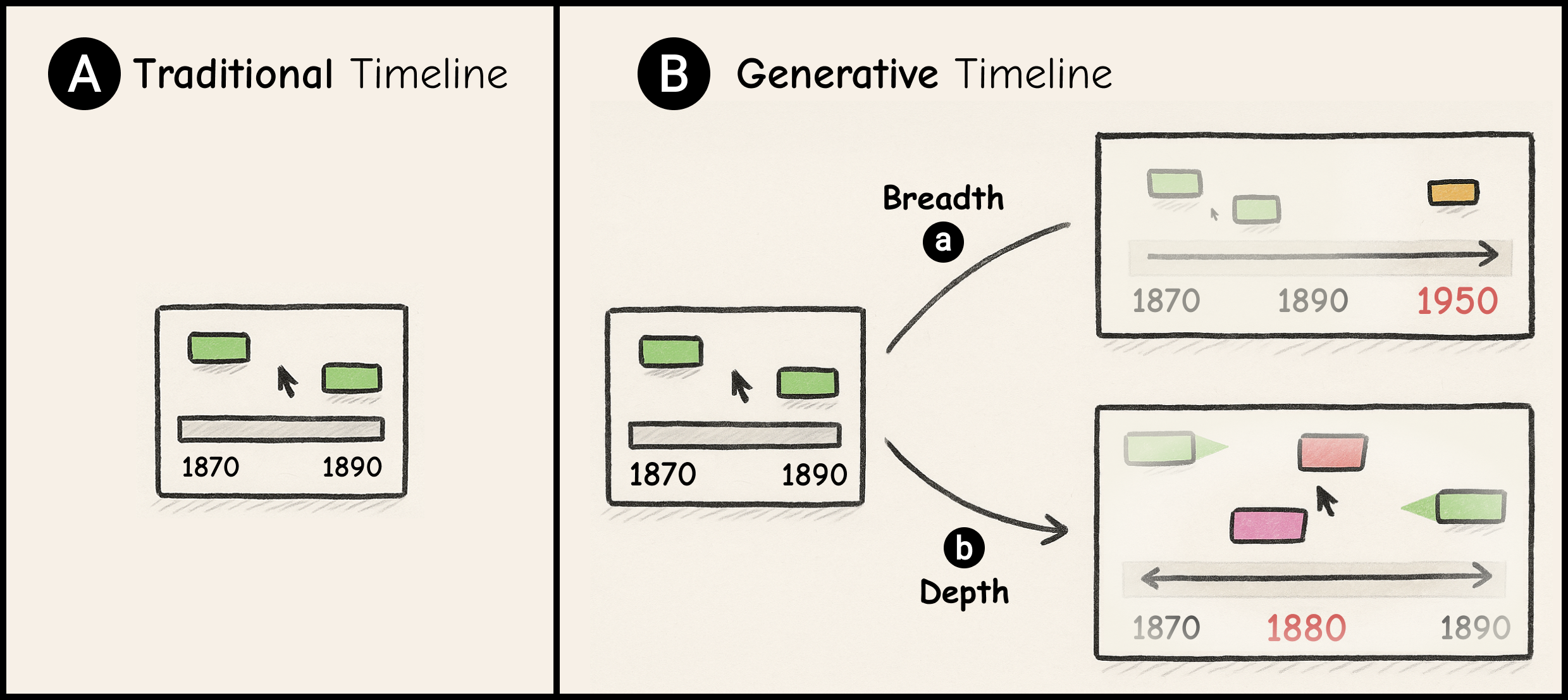}
    \caption{(A) Traditionally, timelines have been constructed manually by experts or automatically from pre-existing documents and data~\cite{shahaf2013information, yildirim2018analysis, akinola2020virtual, remolar2021learning, fulda2015timelinecurator}. As a result, they have primarily functioned as static presentation artifacts, communicating pre-defined narratives with fixed scope and structure, rather than to support open-ended exploration. Recent advances in generative AI open new opportunities to reimagine timelines as interactive tools for flexible exploration and sensemaking. We explore this idea of a generative timeline—a form of timeline that can expand or contract through interaction to support both (a) breadth-based and (b) depth-based exploration—by concretizing it in an interactive system, and conduct a user study to identify various patterns of exploration, discovery, and sensemaking.}
	\Description[short description]{This image shows visually the way timeline expands in depth and in breadth.}
    \label{fig:framework}
\end{teaserfigure}



\maketitle

\section{Introduction}

In an era where generative AI is transforming how people search, learn, and create, interactive systems are beginning to shift from static information presentation toward more dynamic, exploratory experiences~\cite{suh2023sensecape, suh2024dynamic, cao2025generative, min2025malleable, zheng2024disciplink, ma2025garden}. While many of these systems emphasize faster access to information or creative generation~\cite{brade2023promptify, suh2025storyensemble, almeda2024prompting, suh2024luminate, jiang2025orca}, a parallel challenge—and an underexplored opportunity—lies in helping users understand how knowledge itself is formed—how ideas, explanations, and fields emerge, evolve, interact, and accumulate over time. 

For making sense of such developments, timelines have long served as powerful cognitive artifacts for organizing information temporally, helping people identify sequences, patterns, and context from which certain ideas and events emerged~\cite{fulda2015timelinecurator, brehmer2016timelines}. 
However, traditional timelines typically present a fixed selection of events, offering only a partial view of a topic and limited support for flexible exploration. 
Even when certain ideas and events present in a given timeline spark curiosity to explore further, users are often unable to pursue emerging questions beyond the predefined set of events. As a result, timelines are ill-suited for creative, open-ended exploration of knowledge formation over time — such as understanding how ideas originate, branch, gain support, and are reframed across historical and conceptual contexts. 

In the meantime, recent advances in generative AI, particularly large language models (LLMs), have enabled a growing body of work on exploratory interfaces that support open-ended inquiry and sensemaking~\cite{brade2023promptify, suh2023sensecape, jiang2023graphologue, lee2024paperweaver, suh2025storyensemble, zamfirescu2025beyond, peng2025navigating}. While these systems explore diverse interaction styles and representations, timelines—despite being a fundamental medium for temporal reasoning—have received limited attention as generative, LLM-powered interfaces. This raises several questions: (1) what patterns of exploration emerge when a timeline can grow and adapt through interaction? (2) how do specific generative affordances shape users' ability to build and connect temporal knowledge? (3) how does this differ from—and what does it offer beyond—conversational AI interfaces?


To investigate these questions, we developed \textbf{KnowledgeTrail}, an interactive system that instantiates the concept of a generative timeline. KnowledgeTrail supports not only topical exploration, but also reflection on how knowledge forms and evolves through events, ideas, contexts, and relationships over time. Users can request additional events or contextual detail, adjust temporal scope, and surface relationships between events, while the timeline continuously updates in response. In doing so, KnowledgeTrail treats the timeline not as a fixed container of information, but as an evolving representational substrate that responds to user inquiry.
We evaluated KnowledgeTrail through an exploratory study (N=12) in which participants examined topics spanning multiple domains, including \texttt{history} (e.g., \textsf{age of exploration}), \texttt{science} (e.g., \textsf{thermodynamics}), \texttt{computer science} (e.g., \textsf{augmented and virtual reality research}), \texttt{philosophy} (e.g., \textsf{metaphysics}), \texttt{biology} (e.g., \textsf{mycology}), and \texttt{politics} (e.g., \textsf{female world leaders}). These topics provided diverse contexts for observing interaction with evolving temporal representations and for examining how participants reasoned about the development and structure of knowledge within a domain, rather than treating events and ideas as isolated facts. In summary, our work makes the following contributions:

\begin{itemize}

    \item \textbf{KnowledgeTrail}, an interactive system that instantiates generative timelines as a sensemaking medium, illustrating how users can iteratively explore, refine, and relate temporally structured information through direct manipulation and prompting.

    \item \textbf{Empirical findings} characterizing how people engage with generative temporal representations, including five recurring temporal navigation patterns (\emph{\textbf{\hlcb{TeleportationPattern}{temporal teleportation}}}, \emph{\textbf{\hlcb{ProgressivePattern}{progressive}}} \emph{\textbf{\hlcb{ProgressivePattern}{zoom-in}}}, \emph{\textbf{\hlcb{OscillationPattern}{local oscillation}}}, \emph{\textbf{\hlcb{ConvergencePattern}{convergence-then-broadening}}}, and \emph{\textbf{\hlcb{MultiEraPattern}{complex multi-era navigation}}}), instances of serendipitous discovery, and relational sensemaking behaviors that differ in frequency and character from conversational AI interfaces.

    
\end{itemize}

\begin{figure*}[h!]
    \centering
  \includegraphics[trim=0cm 0cm 0cm 0cm, clip=true, width=0.95\textwidth]{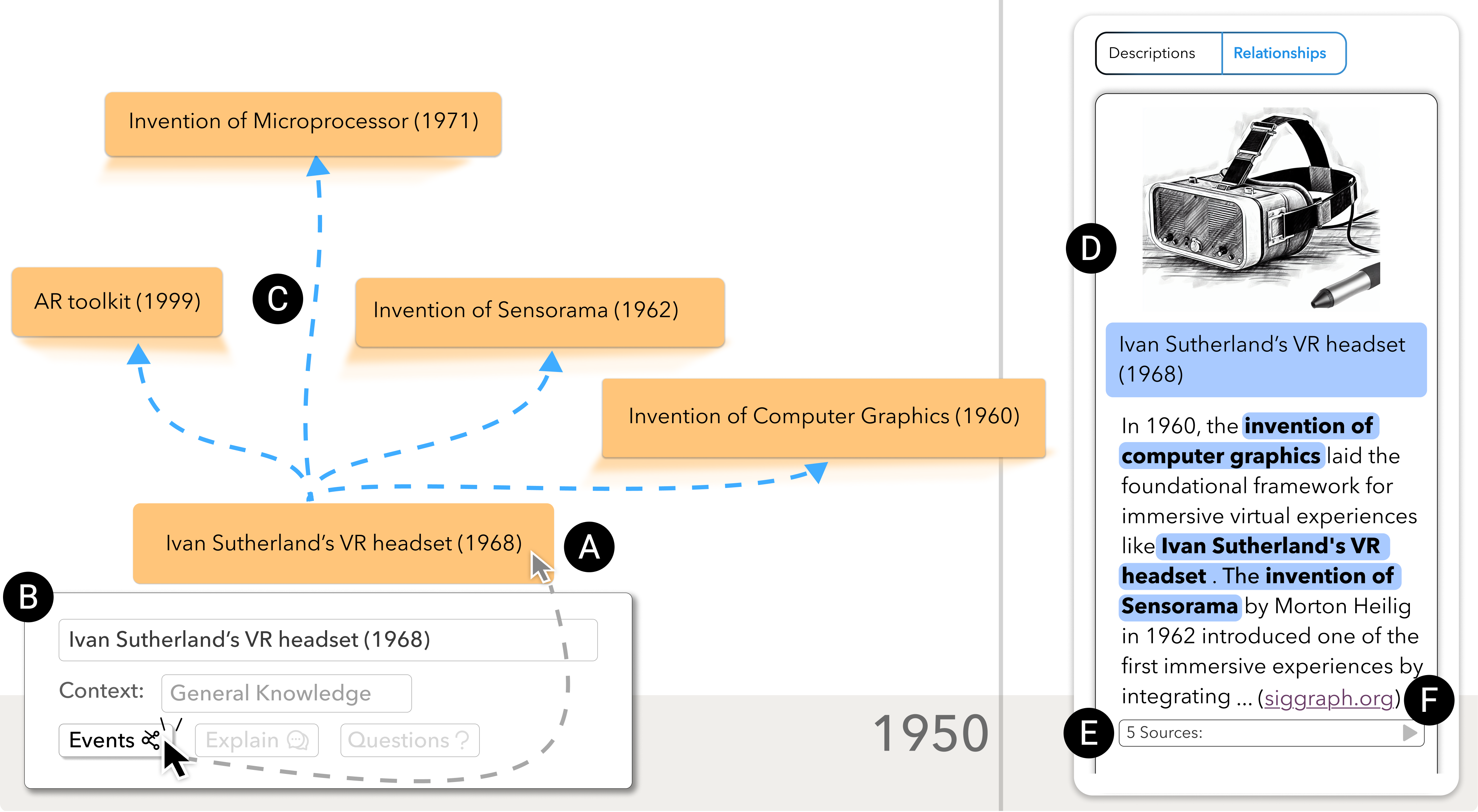}
  \caption{An example of \textsc{event generation} in KnowledgeTrail: Users can generate events related to a topic by (A) hovering over a topic on timeline and (B) selecting the \hlcb{button}{Events} button in the \textsf{Expand Bar} or with the \textsf{Search Bar} (Fig.~\ref{fig:ContextualEventGeneration}). The (C) arrows trace the path of generation to help track the exploration path. A description on how the events are related is then (D) displayed in the \textsf{Side Panel}. (E) Users are presented with (E \& F) links to sources. They can select (E) citation box or (F) inline link to open a new tab in the browser.}
  \Description{This is an image of the KnowledgeTrail interface, featuring timeline view, search box, and side panel. The search box, located at the top middle, is where users can type in a prompt. The timeline view shows event nodes placed based on the year it happened. The side panel on the right side shows information about a searched topic generated by AI.}
	\label{fig:event-generation}
\end{figure*}

\section{Related Work}

\subsection{Timelines as Sensemaking Representations}

The timeline representation was first introduced by Joseph Priestley in his 1765 book \textit{A Chart of Biography}, enabling a departure from purely textual formats that make it difficult to discern relationships between ideas and events distributed across a narrative. By arranging elements sequentially and spatially, timelines serve as a powerful cognitive artifact for identifying temporal order, comparing events and periods, and reasoning about change over time~\cite{10.1145/504216.504245, grossman2010chronicle, burch2008timeline, plaisant1996lifelines, nguyen2014schemaline, 1249028, bui2007timeline, zhao2012timeslice}. Beyond organizing events, timelines have also been used—implicitly or explicitly—to support understanding of how ideas, practices, and bodies of knowledge develop over time, by making temporal relationships visible.

This affordance has led to the adoption of timelines across a wide range of domains, including history~\cite{bennett2012chronotape}, journalism~\cite{segel2010narrative, chen2023does}, management~\cite{aseniero2020activity}, and data analysis~\cite{andre2007continuum, huang2025narrative}, where they serve as effective tools for facilitating sensemaking and contextual understanding. Building on these foundations, researchers have developed interactive timeline systems that allow users to navigate and interrogate temporal data. For example, Chronicle~\cite{grossman2010chronicle} uses timelines to explore graphical document workflows, Lifelines~\cite{plaisant1996lifelines} provides interactive visualizations of personal life events, and SchemaLine~\cite{nguyen2014schemaline} employs timelines to enhance sensemaking by integrating schematic structure with temporal progression. Using techniques such as filtering, zooming, and linking, these systems demonstrate the expressive power of timelines as sensemaking artifacts. These systems collectively established core affordances of traditional timelines—temporal axis organization, spatial ordering, event stacking, density management, and filtering—that a generative timeline both inherits and extends.


However, traditional timeline systems have presented pre-authored representations in which the set of events, their structure, and their granularity are determined in advance. Even when interaction is supported, it operates over a static underlying model (dataset). While some systems have examined interactive timeline navigation in specific contexts~\cite{plaisant1996lifelines, jensen2003visualizing, zgraggen2015s}, generative timelines—where the structure itself adapts dynamically (Fig.~\ref{fig:framework}) to user inquiry rather than operating over a fixed dataset—have received limited attention. As a result, how users navigate and make sense of knowledge when the timeline itself is open-ended and generative remains an open question.


\begin{figure*}[h!]
	\centering
	\includegraphics[width=\textwidth]{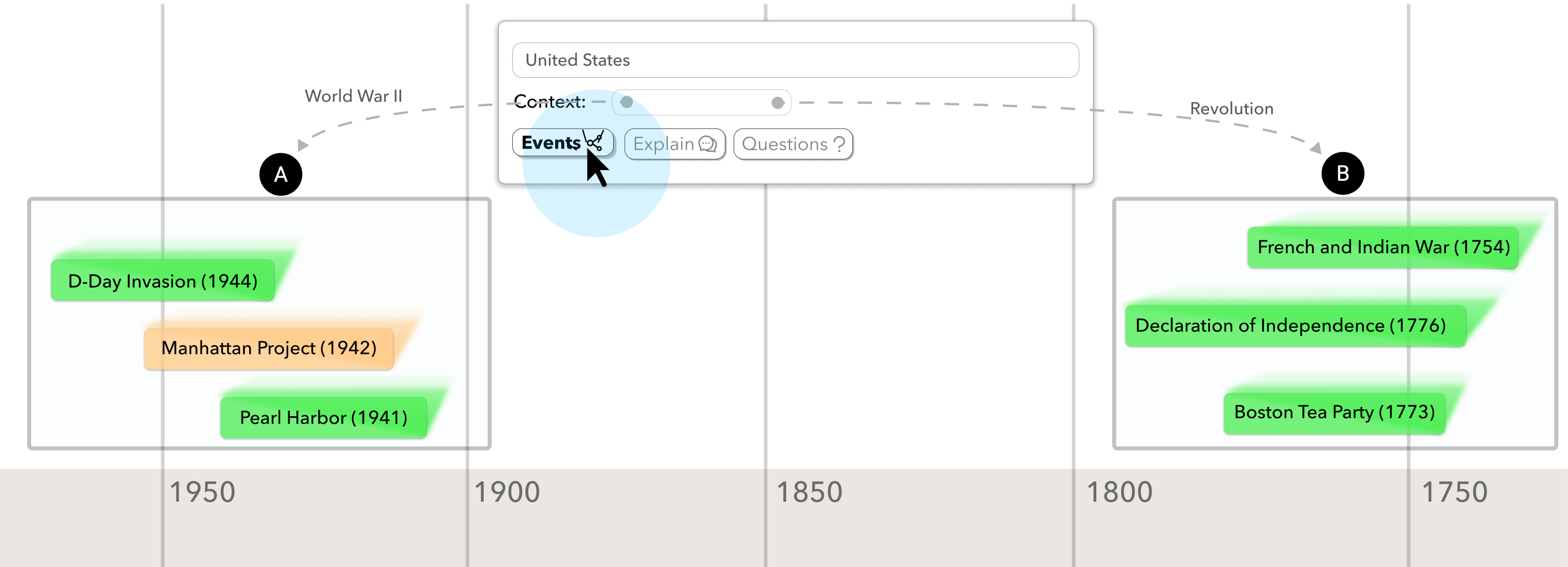}
	\caption{\textit{Contextual} \textsc{Event Generation}: Users can steer the type of event generation within a single prompt (e.g. United States). (A) One context (\textit{World War II}) can generate events focusing around that topic (e.g., \textit{Pearl Harbor}), while (B) another context (\textit{Revolution}) can help constrain the generation to a completely separate group of events (e.g., \textit{Boston Tea Party}) when the \textsf{\hlcb{button}{Events}} button is clicked.}
	\Description[short description]{This image shows how users can steer what type of events they want to see. For example, in this figure, it shows United States as the prompt in the search box. It shows two groups of events: one generated from World War II as context and the other from Revolution as the other context.}
	\label{fig:ContextualEventGeneration}
\end{figure*}

\subsection{Exploratory Sensemaking and External Representations}

Research on exploratory sensemaking emphasizes that understanding often emerges through iterative cycles of questioning, externalization, and revision~\cite{russell1993cost, chang2019searchlens}. External representations such as diagrams, maps, and timelines play a central role in this process by enabling users to offload cognition, compare alternatives, and revisit prior interpretations~\cite{kirsh2010thinking}. Prior work has shown that representations that support flexible restructuring can significantly shape how people reason about complex information spaces~\cite{suh2023sensecape}. In many domains, this restructuring involves learning not only new facts, but also how explanations, concepts, or interpretations emerge, evolve, and relate to one another over time~\cite{kang2022threddy, ma2025garden}. This iterative externalization-revision cycle~\cite{russell1993cost} suggests that effective sensemaking tools should support not just consumption of information but active restructuring of it—a principle that motivates KnowledgeTrail's design, in which users can continuously expand, contract, and reorganize the temporal representation in response to emerging understanding.

KnowledgeTrail builds on this perspective by treating timelines not as static summaries, but as interactive substrates for sensemaking, where users can continually reshape temporal structure as their inquiry unfolds. Rather than assuming a fixed narrative, the system supports iterative exploration in which temporal organization itself becomes part of the reasoning process.

\subsection{Exploratory Interfaces Powered by Generative AI}

Recent work has explored the integration of generative AI into interactive systems, enabling users to produce content, explanations, and alternatives on demand~\cite{suh2023sensecape, suh2024luminate, brade2023promptify, zamfirescu2025beyond, cao2025compositional}. While much of this work focuses on conversational interfaces or generative output as end results, fewer systems examine how generative models can participate in the ongoing construction of interactive representations. Prior systems have introduced novel interaction techniques and representational forms to scaffold exploration and abstraction, yet they often treat generated content as something to be consumed rather than structurally reorganized.

In KnowledgeTrail, generative AI is not used merely to populate a visualization, but to support continuous restructuring of temporal representations in response to user interaction. This raises distinct interaction design challenges, including how to maintain legibility, support user agency, and balance serendipity with control when representations themselves are mutable. By making temporal structure dynamic, this work highlights how users reason about the emergence and evolution of knowledge, rather than treating timelines as static summaries of pre-assigned narratives.

Unlike prior timeline systems that operate over fixed corpora — where a bounded, pre-authored set of events is laid out and navigated — KnowledgeTrail operates over an unbounded information space in which events do not exist until queried and are synthesized rather than retrieved. This is not merely a difference of scale: it changes the nature of exploration from navigating `\textbf{known unknowns}' (finding events the user suspects exist) to encountering `\textbf{unknown unknowns}' (surfacing events and relationships the user did not know to ask for). KnowledgeTrail explores what happens when the structure of a timeline itself becomes generative and adaptive through interaction, contributing an interaction paradigm and design insights applicable across domains that involve temporal sensemaking.


\begin{figure}[h!]
	\centering
	\includegraphics[trim=2cm 0cm 0cm 0cm, clip=true, width=0.48\textwidth]{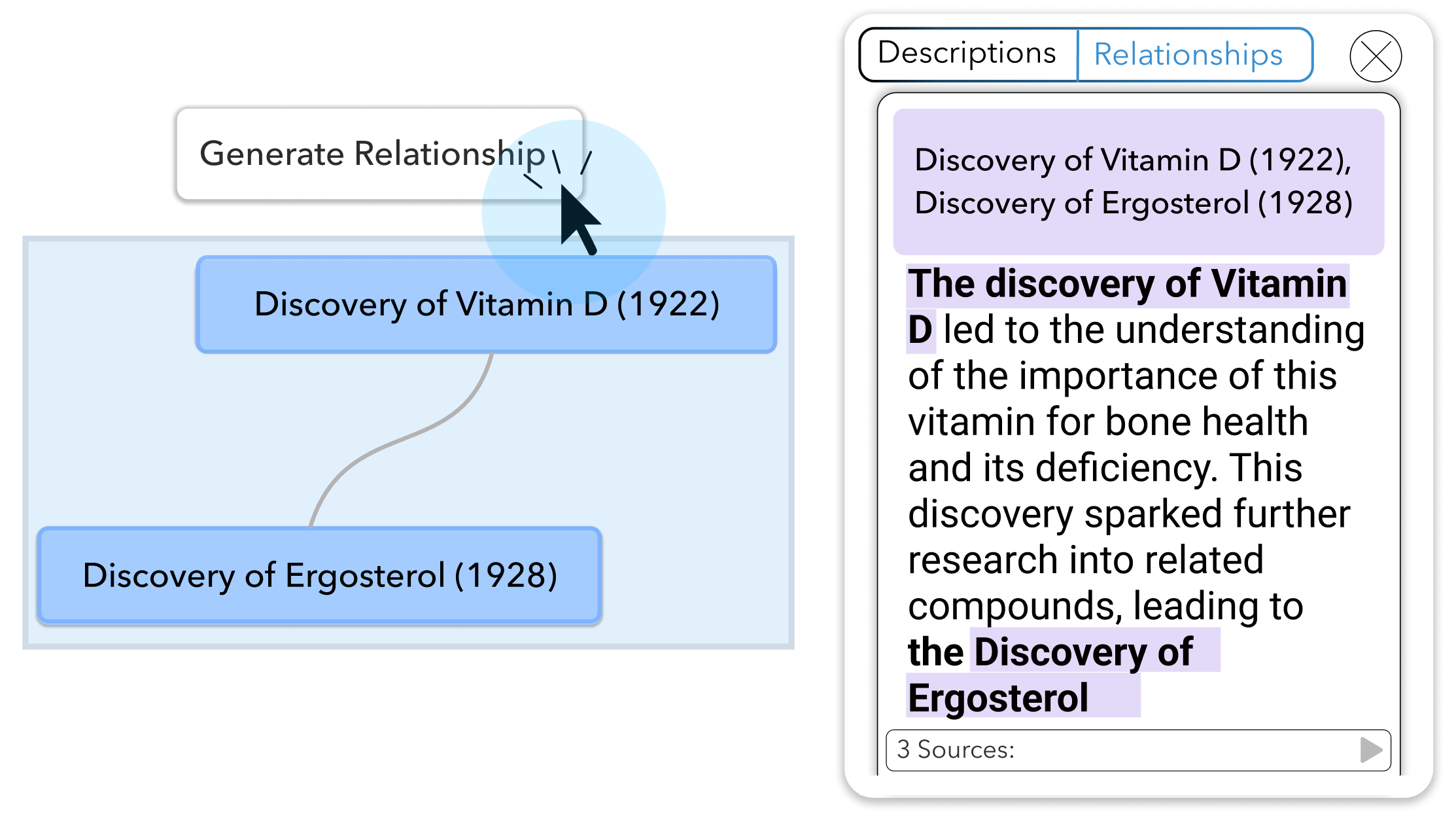}
	\caption{Generating Relationships: Users can explore the relation between two or more separate events generated on the timeline by selecting the topics and clicking the \textsf{\hlcb{button}{Generate Relationship}} button. A paragraph describing the potential relationship between the events is then generated and displayed in the \textsf{Side Panel} under the \textcolor{ACMBlue}{\textsf{Relationships}} tab.}
	\Description[short description]{This image shows a feature in KnowledgeTrail that allows users to ask and learn about the selected events or topics. It shows two events connected by an edge: Discovery of Vitamin D (1922) and Discovery of Ergostrol (1928). On the right, it shows a side panel with the description of how they are related.}
	\label{fig:generate-relationship}
\end{figure}

\section{\sys{}}

\subsection{User Interface \& Interactions}
\label{sec:interface}

KnowledgeTrail is a web-based interactive system built on ReactFlow. Its interface centers on a \textsf{Timeline View} (Figs.~\ref{fig:event-generation} \& \ref{fig:ContextualEventGeneration}) — an infinite canvas on which generated event nodes are placed chronologically. A dynamic timeline bar at the bottom automatically expands or contracts to fit the temporal range of events currently on the canvas. Users can pan, zoom, drag, and delete nodes freely. All content generation is powered by OpenAI's GPT-4; prompts for each interaction are detailed in Table~\ref{table:llm-prompts} in the Appendix.

\begin{figure*}[hbt!]
	\centering
	\includegraphics[trim=3cm 50cm 20cm 0cm, clip=true, width=0.8\textwidth]{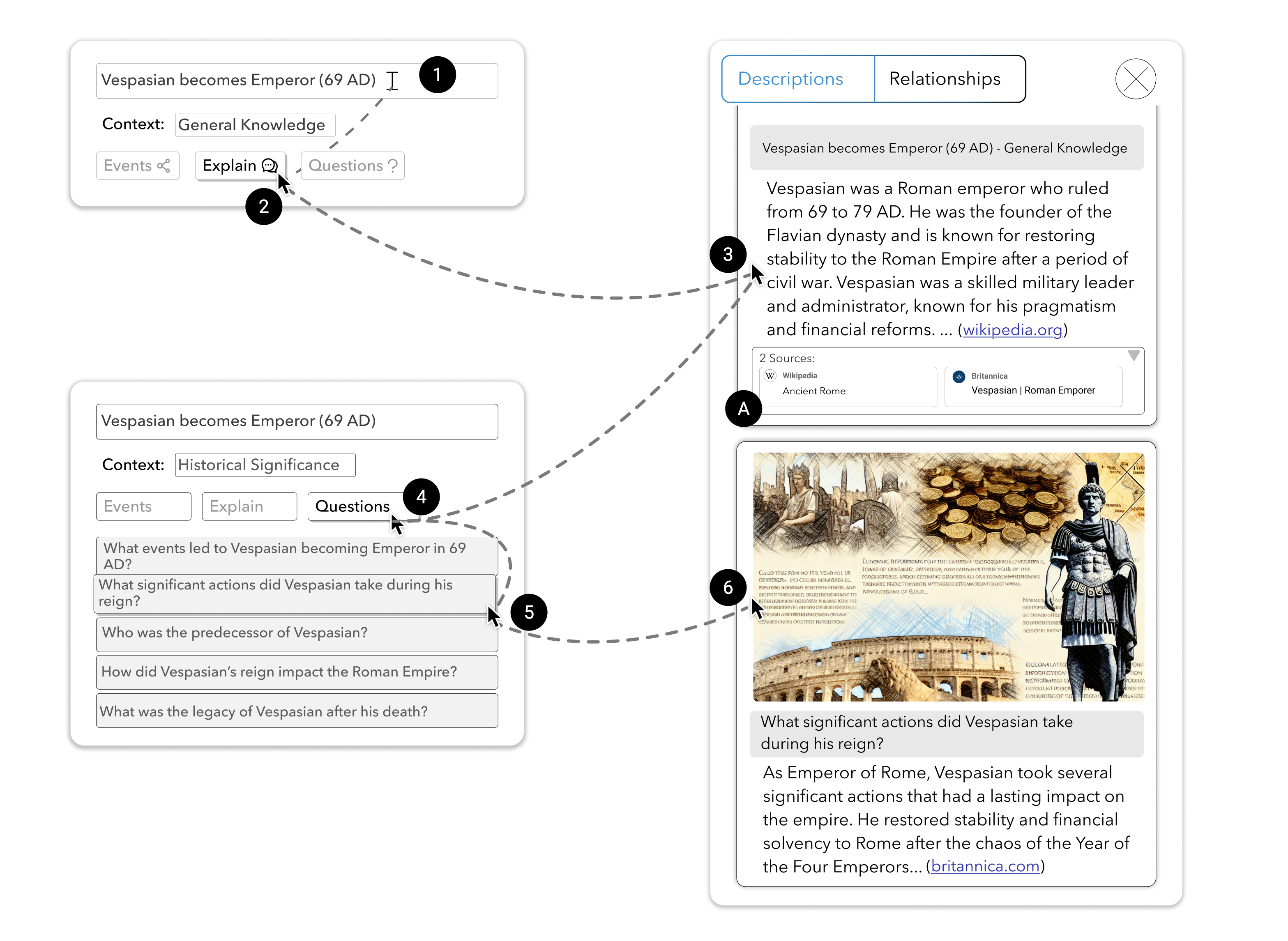}
	\caption{Generating Descriptions Flow: Users can (1) learn more details about a topic in the \textsf{Search Box} or event in the timeline by (2) using the \textsf{\hlcb{button}{Explain}} button. A detailed explanation is then (3) generated in the \textsf{Side Panel} for the given topic/event. To facilitate further exploration on the topic, the \texttt{\hlcb{button}{Questions}} button (4) can be clicked to generate a list of questions relating to the topic and context. Users can (5) select one of these questions to learn more about and a response will be (6) generated again on the \textsf{Side Panel}. (A) Source citations are available for users to verify the accuracy of the information.}
	\Description[short description]{This image shows a workflow users may take, while showcasing how one can use the \textsc{Explain} and \textsc{Questions} buttons. It shows that it generates an explanation of the event 'Vespasian becomes Emperor (69 AD)' and how the user selects the \textsc{Questions} button to generate a list of questions and then clicks one of the questions to generate its explanation.}
	\label{fig:explain-and-questions}
\end{figure*}

\textbf{Search Box (Fig.~\ref{fig:ContextualEventGeneration})}
Users initiate exploration by entering a topic and optional context into the \textsf{Search Box}. Clicking \events generates a set of nodes, color-coded by event type (\pill{LegendTheory} Theory, \pill{LegendDiscovery} Discovery, \pill{LegendInvention} Invention, \pill{LegendPolitics} Politics, \pill{LegendArt} Art, \pill{LegendEconomics} Economics, \pill{LegendOther} Other), places them onto the timeline, and opens the \textsf{Side Panel} with a paragraph explaining the relationships between generated events (Fig.~\ref{fig:event-generation}). Clicking \explain generates a detailed description of the topic alongside an AI-generated image. Clicking \questions produces five follow-up questions that can be selected to trigger further generation, supporting continuous exploration (Fig.~\ref{fig:explain-and-questions}).

\textbf{Expand Bar (Fig.~\ref{fig:event-generation}B)} 
Hovering over any event node reveals the \textsf{Expand Bar}, which mirrors the \textsf{Search Box} interface with the event's name pre-filled. The \events button generates related events with arrows indicating their origin; \explain generates a description of the event in the \textsf{Side Panel}; and \questions produces related follow-up questions. A short 1–2 sentence summary is also generated inline within the \textsf{Expand Bar} to help users quickly orient themselves.

\begin{figure}[h!]
	\centering
        \includegraphics[trim=0cm 0cm 0cm 0cm, clip=true, width=0.48\textwidth]{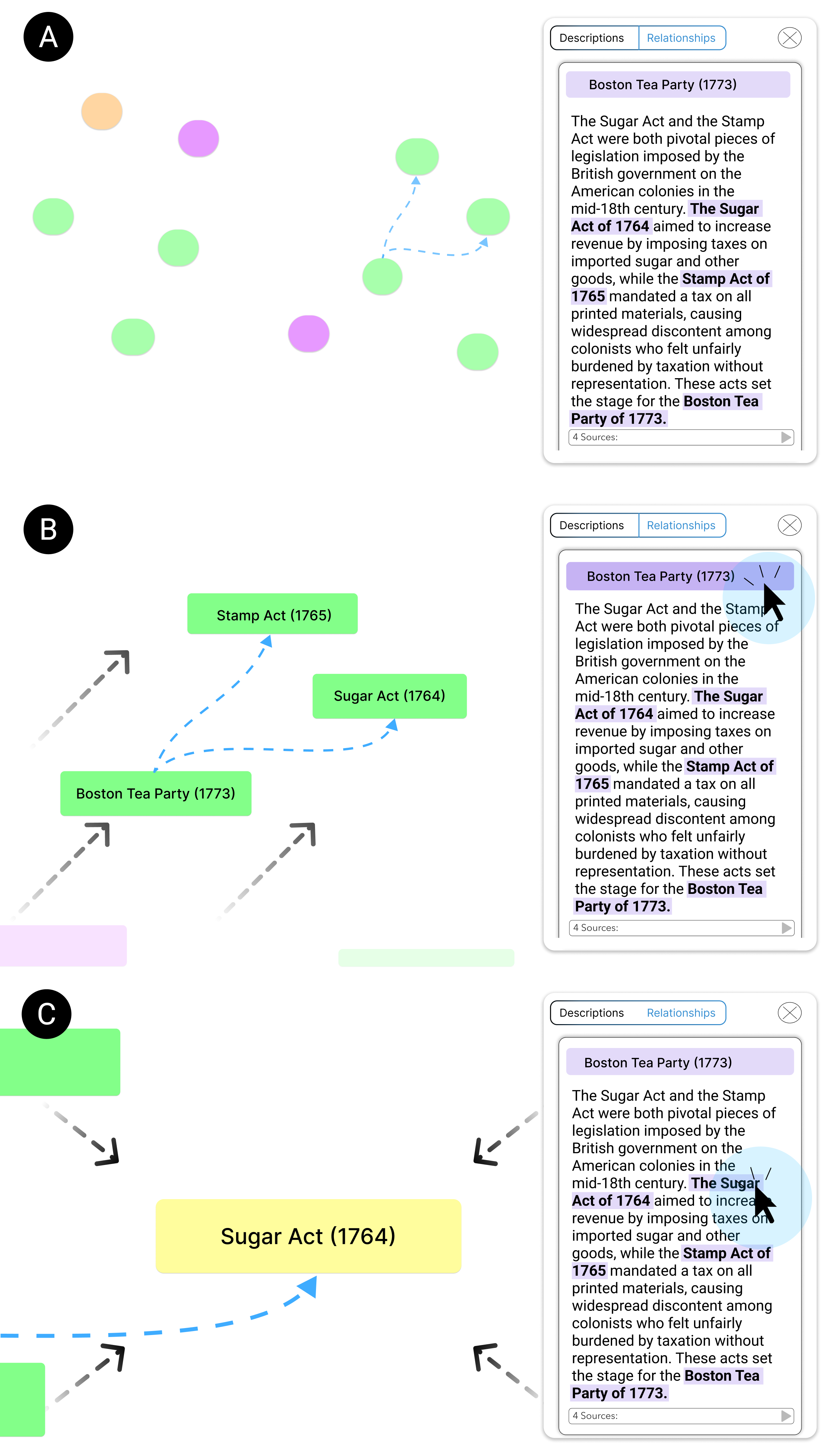}
	\caption{Navigation from \textsf{Side Panel}: Users can return to previously generated content by clicking a result title in the \textcolor{ACMBlue}{\textsf{Relationships}} tab (B), which adjusts the \textsf{Timeline View} to show all referenced events. Selecting an event (C) zooms into that point in the timeline. Hovering over an event in the \textsf{Side Panel} highlights it on the timeline for easy identification.}
	\Description[short description]{This image shows how users can navigate to a select group of nodes generated from prompt. It shows how users can select a title in the side panel to move the view to the group of nodes generated from that title. It also shows that users can select an event listed in the description to move the view to focus on that selected event node.}
	\label{fig:nav-side-panel}
\end{figure}

\textbf{Side \& Legend Panel (Figs.~\ref{fig:event-generation}D \& \ref{fig:nav-side-panel})} 
The \textsf{Side Panel} displays generated relationship paragraphs (under a \tab{TabColor}{\textsf{Relationships}} tab) and detailed descriptions or question responses (under a \tab{TabColor}{\textsf{Descriptions}} tab). Events are bolded and hyperlinked for direct navigation (Fig.~\ref{fig:nav-side-panel}). The \textsf{Legend Panel} displays color-coded event types and doubles as a filter: clicking a type highlights matching nodes and adjusts the view to focus on them.

\textbf{Generate Relationship (Fig.~\ref{fig:generate-relationship})} 
Selecting two or more nodes surfaces a \relationships button, which populates the Side Panel with a paragraph describing the potential connections between the selected events and renders a line between them on the canvas.

\textbf{Semantic Zoom (Fig.~\ref{fig:nav-side-panel})} 
As the canvas grows dense, zooming out past a threshold collapses nodes into color-coded dots, allowing users to identify temporal clusters and gaps at a glance. Hovering over a collapsed dot displays the event name and any previously generated summary.

\textbf{Source Citation UI (Figs.~\ref{fig:event-generation}E-F \& \ref{fig:explain-and-questions}A)} 
To support evaluation of generated content, inline citation links appear next to key claims in the \textsf{Side Panel}, and a collapsible \textsf{Citation Panel} reveals the associated sources.

\begin{figure}[h!]
    \centering
    \begin{subfigure}[t]{0.47\textwidth}
    \includegraphics[trim=0cm 0cm 0cm 0cm, clip=true, width=1\columnwidth]{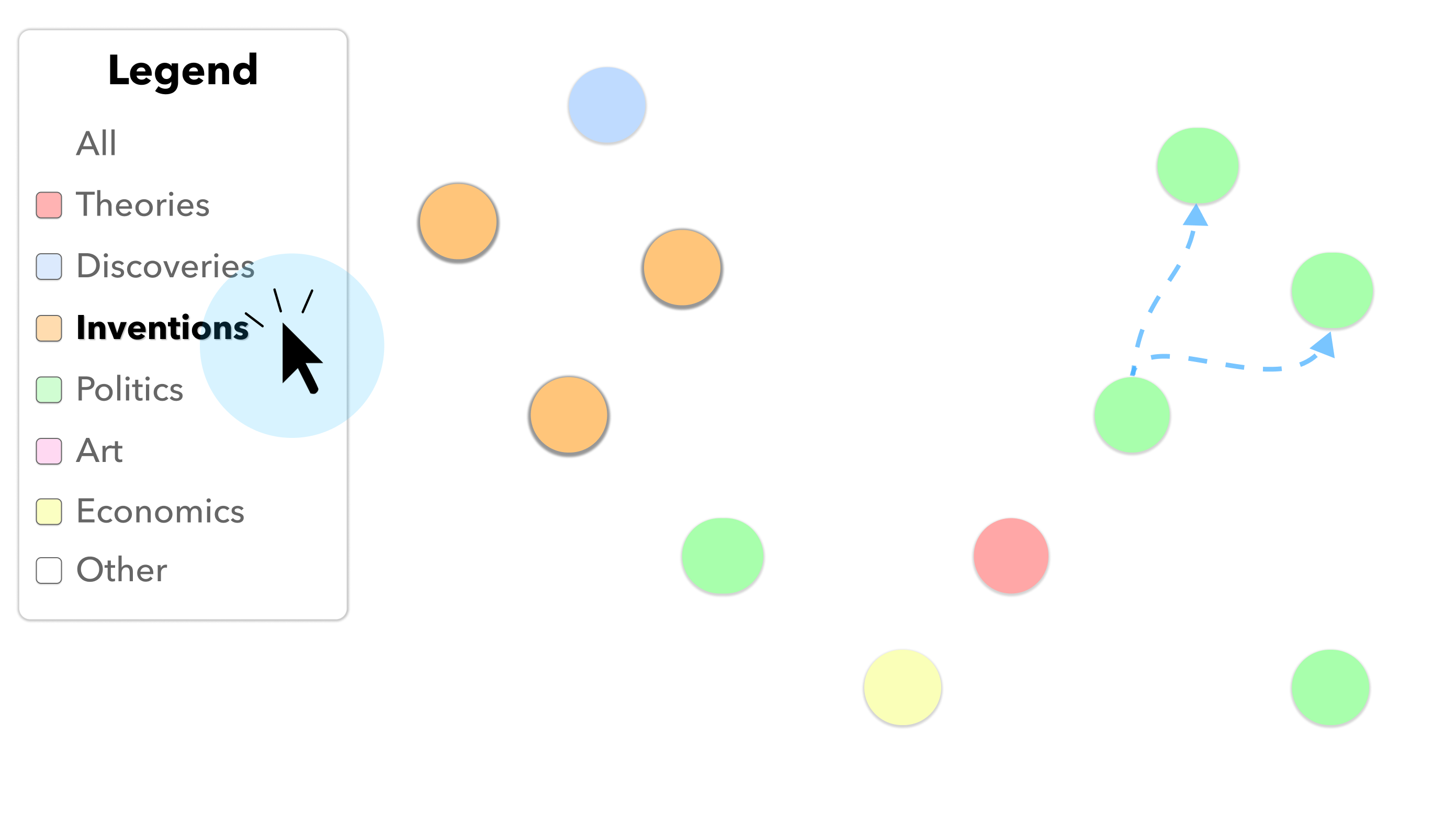}
    \caption{No Event Types selected}
    \label{fig:search-keywords}
    \end{subfigure}
    \begin{subfigure}[t]{0.47\textwidth}
        \includegraphics[trim=0cm 0cm 0cm 0cm, clip=true, width=\textwidth]{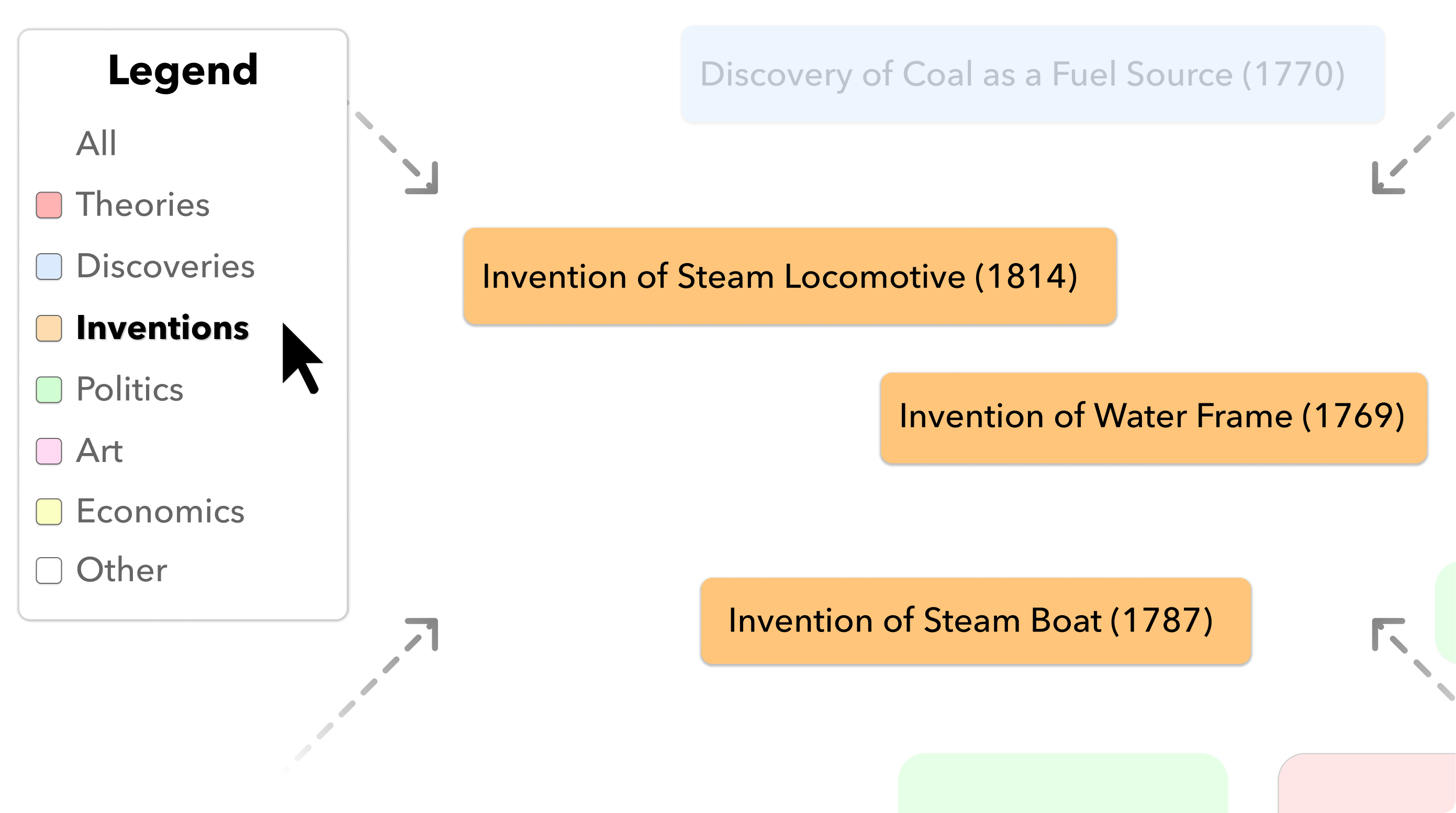}
        \caption{Selected the Event Type \pill{LegendInvention} \textsf{Inventions}}
        \label{fig:search-filter}
    \end{subfigure}
    \caption{Navigation from \textsf{Legend Panel}: Users can filter and efficiently navigate (b) to a select group of nodes corresponding to the label they select. Matching nodes of that type (e.g., \textsf{Inventions}) are highlighted and the view zooms in (a -> b) to focus on these events.}
    \Description[short description]{This image shows a legend panel located at the top left of the KnowledgeTrail interface. The legends feature different types of events\: theories, discoveries, inventions, politics, art, economics, and other. It shows how users can select the label and have the view move to focus on a group of events that correspond to the selected label.}
    \label{fig:legend}
\end{figure}

\section{Exploratory Study}
\label{sec:study1}


The goal of our study was to understand how generative timelines shape exploratory sensemaking behaviors and how timeline-based interaction compares to a conversational baseline for open-ended knowledge exploration. Rather than testing predefined hypotheses, we aimed to surface (1) patterns of interaction, (2) sensemaking strategies, and (3) discoveries that emerge when temporal structure can be dynamically expanded and reorganized through interaction. 
To foreground generative timeline interactions and reduce verification-related cognitive load (which we observed in our pilot studies), participants used a version of KnowledgeTrail without citation features.
The study was approved by the university's ethics board and targeted the following research questions:

\begin{itemize}
    \item \textbf{RQ1.} How do users explore topics using a generative timeline, and how do their exploration patterns differ from conversational interfaces?
    \item \textbf{RQ2.} How do specific generative-timeline features (e.g., contextual expansion, relationship prompting, semantic zoom, navigation) shape users' ability to build, revisit, and connect temporal explanations?
    \item \textbf{RQ3.} What roles do users envision for generative timelines in learning and knowledge work, and what concerns do they raise around credibility and steering?
\end{itemize}

\noindent\textbf{Participants}. We recruited 12 participants (6F, 6M; M$_{age}$=27.6 [22,44], SD$_{age}$=5.8) from a local R1 university and Upwork. All (P1-P12) had prior experience using ChatGPT for learning; most used AI tools at least daily (9 participants) and reported relying on them to a moderate or high extent. Full demographics are in Table~\ref{tab:participant_info} in the Appendix.



\noindent\textbf{Study Design and Tasks}. 
After a pre-study survey on demographics and AI tool experience, 
participants used both systems across three topics total: two in KnowledgeTrail (one assigned — either ROME or EXPLORATION — and one self-selected) and one in ChatGPT (the remaining assigned topic). Assigned topics and systems were counterbalanced across participants to control for order effects. For KnowledgeTrail, we first provided a tutorial covering each feature. 
They were instructed to imagine preparing a group presentation on each topic (see Table~\ref{table:user-study-topic} in Appendix) and asked to think aloud throughout. Sessions concluded with a post-study survey and semi-structured interview. Participants received \$30 for this 1.5-hour study.

\noindent\textbf{Baseline.} ChatGPT was chosen as a baseline to isolate the effect of system design: both share the same underlying LLM, allowing us to attribute behavioral differences to KnowledgeTrail's interaction paradigm rather than model capability.


\noindent\textbf{Measures and Analysis}. To assess how KnowledgeTrail supports exploratory sensemaking, we used a combination of quantitative and qualitative measures. We administered the Creativity Support Index (CSI)~\cite{carroll2009creativity} to capture perceived support for exploration, expressiveness, enjoyment, immersion, and results worth effort — chosen because the study focused on curiosity-driven exploration rather than task efficiency, and because CSI has been widely used to evaluate open-ended exploratory systems~\cite{suh2023sensecape, suh2025storyensemble}. We also collected Likert-scale ratings on the perceived usefulness of key interface components, and asked participants to compare their experience with ChatGPT across dimensions of exploration and sensemaking. Finally, we conducted thematic analysis of think-aloud protocols, screen recordings, and post-study interviews to identify recurring exploration strategies, moments of surprise or insight, and how participants constructed narratives and connections across events.

\section{Results}
\label{sec:study1_results}

Participants found KnowledgeTrail usable (Easy to Use: 8 Strongly Agree, 3 Agree, 1 Neutral; Easy to Learn: 9 Strongly Agree, 3 Agree) and preferred it over ChatGPT for both exploring (8 Strongly Agree, 4 Agree) and making sense of topics (7 Strongly Agree, 2 Agree, 3 Neutral). Overall support for curiosity-driven exploration was high, reflected in a CSI score of 85.94 (SD=4.32)~\cite{carroll2009creativity}; full factor-level results are in Table~\ref{tab:csi}. A post-hoc accuracy analysis of LLM-generated content found high accuracy for event occurrence (98.3\%), event year (98.1\%), and descriptions (100\%), with more errors in relationship explanations (87\%); More details in Tables~\ref{table:accuracy-analysis}–\ref{table:error-examples} in Appendix.


\begin{figure}
    \centering
    \includegraphics[trim=0cm 0cm 0cm 0cm, clip=true, width=0.9\linewidth]{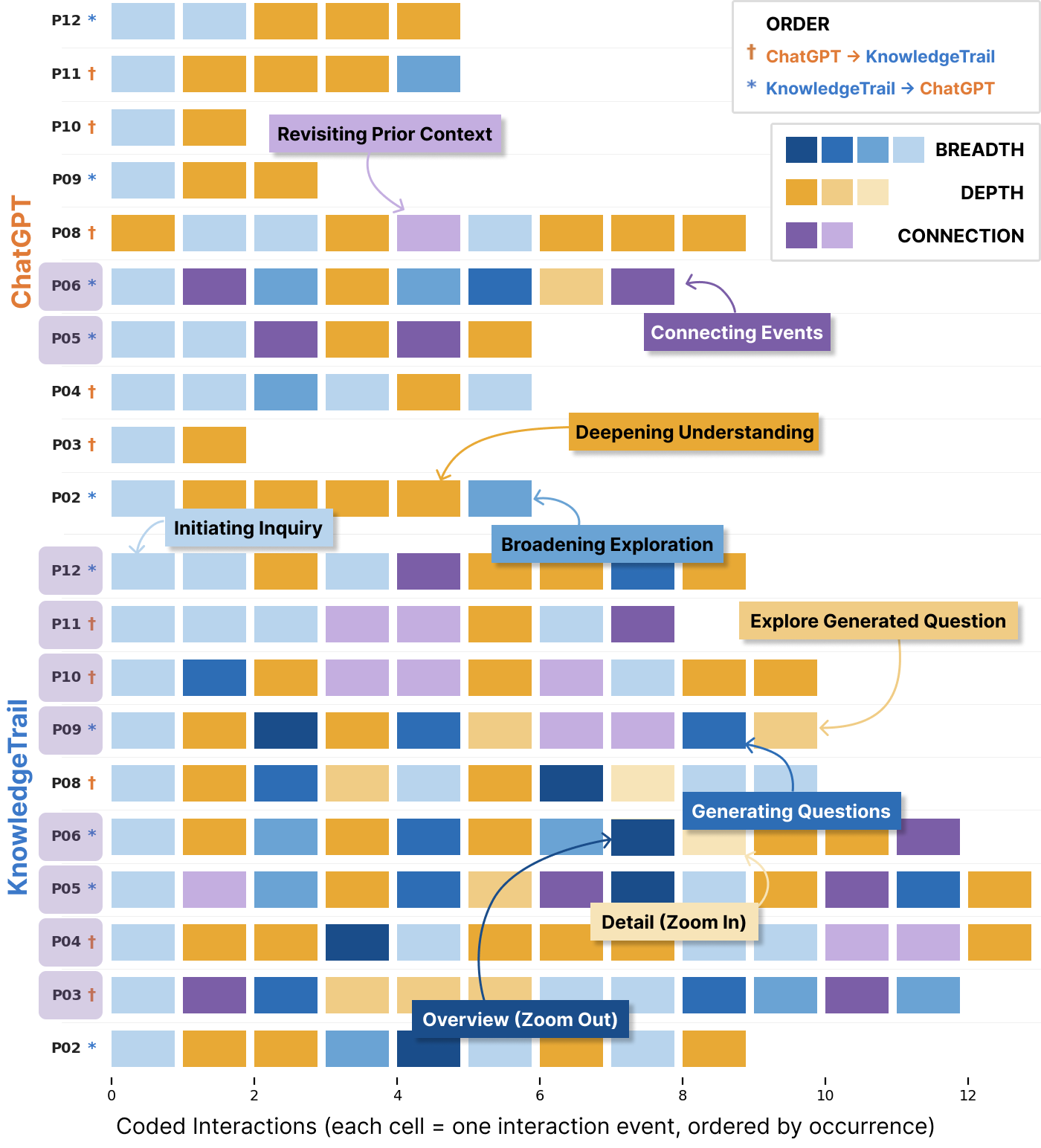}
    \caption{Interaction sequences for ChatGPT (top) and KnowledgeTrail (bottom) across n=10 paired participants (P1, P7 excluded due to missing data; based on assigned topics---ROME and EXPLORATION---to enable direct system comparison). Each cell represents one coded interaction event, ordered by occurrence. 
    Colors indicate exploration dimensions: \raisebox{-0.4pt}{{\includegraphics[scale=0.125]{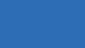}}} Breadth (e.g., \textsf{\hlcb{button}{Events}}), \raisebox{-0.4pt}{{\includegraphics[scale=0.125]{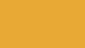}}} Depth (e.g., \textsf{\hlcb{button}{Explain}}), \raisebox{-0.4pt}{{\includegraphics[scale=0.125]{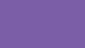}}} Connection (e.g., \textsf{\hlcb{button}{Generate Relationship}});
    See Table~\ref{tab:coding-scheme} for details about coding scheme.
    While \emph{connection}-type interactions (\raisebox{-0.4pt}{{\includegraphics[scale=0.125]{figures/connection-cell.png}}}) appeared in 80\% of KnowledgeTrail sessions, they appeared in only 20\% of ChatGPT sessions. Interestingly, this included \textsf{\hlcb{TeleportationPattern}{P05$^*$}} and \textsf{\hlcb{TeleportationPattern}{P06$^*$}}---both of whom had used KnowledgeTrail first, \textbf{suggesting that KnowledgeTrail's \hlcb{button}{Generate Relationships} feature primed relational thinking and inquiry whereas they did not seem to naturally emerge in conversational interface}.}
    \Description{Two horizontal sequence charts comparing ChatGPT (top) and KnowledgeTrail (bottom) across 10 participants (P02–P12, excluding P01 and P07). Each row represents one participant; each cell is a coded interaction event ordered by occurrence. Cells are color-coded by exploration dimension: green for Breadth, blue for Depth, and purple for Connection. KnowledgeTrail rows are visibly longer and more varied in color, while ChatGPT rows are shorter and predominantly green and blue. Purple Connection cells appear in KnowledgeTrail rows for most participants, but only in the ChatGPT rows of P05 and P06, who are marked with asterisks indicating they used KnowledgeTrail first.}
    \label{fig:interaction_sequence}
\end{figure}

\begin{figure}[h!]
    \centering
  \includegraphics[trim=0cm 0cm 0cm 0cm, clip=true, width=0.48\textwidth]{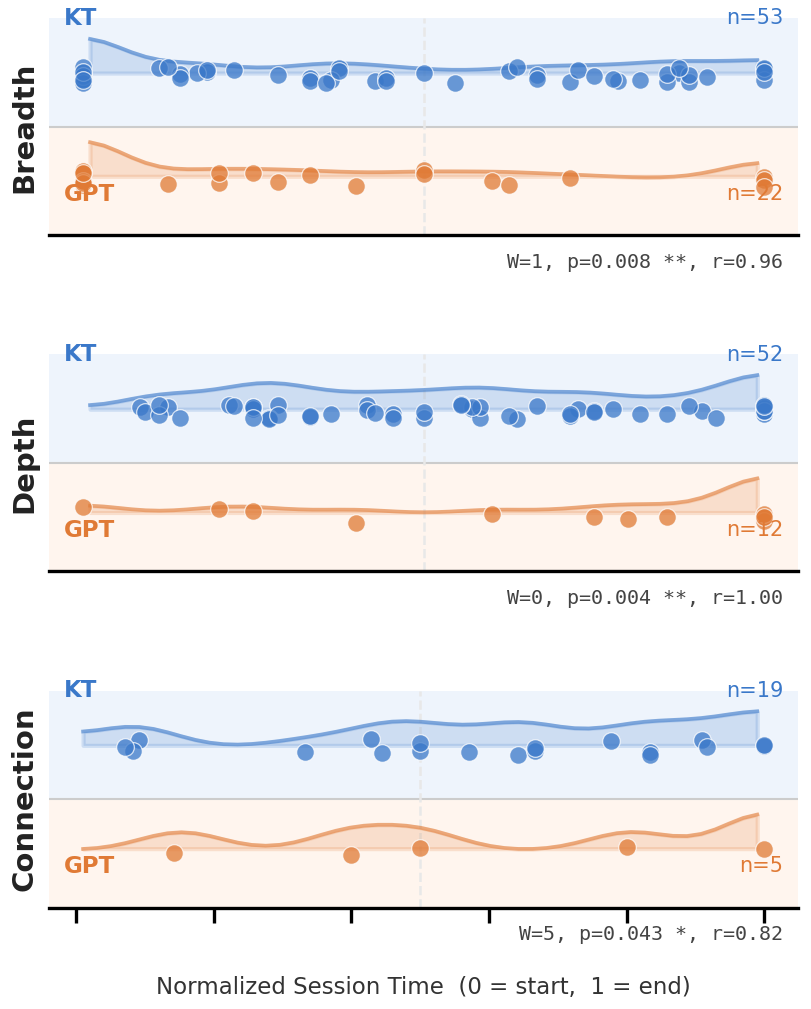}
  \caption{Temporal distribution of interaction events by exploration dimension and total interaction volume, comparing KnowledgeTrail (blue) and ChatGPT (orange) across n=10 participants (within-subjects, fixed session duration). * p<.05, ** p<.01, n.s. = not significant.}
  \Description{Three dot-plot panels comparing KnowledgeTrail (blue dots, top row) and ChatGPT (orange dots, bottom row) across normalized session time (0 = start, 1 = end) for Breadth (n=53 KT, n=22 GPT), Depth (n=52 KT, n=12 GPT), and Connection (n=19 KT, n=5 GPT) interactions. KnowledgeTrail shows substantially more interaction events in all three dimensions, distributed across the session. Statistical annotations indicate significant differences for Breadth (W=1, p=0.008, r=0.96), Depth (W=0, p=0.004, r=1.00), and Connection (W=5, p=0.043, r=0.82).}
	\label{fig:temporal-distribution}
\end{figure}


\begin{figure*}[h!]
    \centering
  \includegraphics[trim=0cm 0cm 0cm 0cm, clip=true, width=0.85\textwidth]{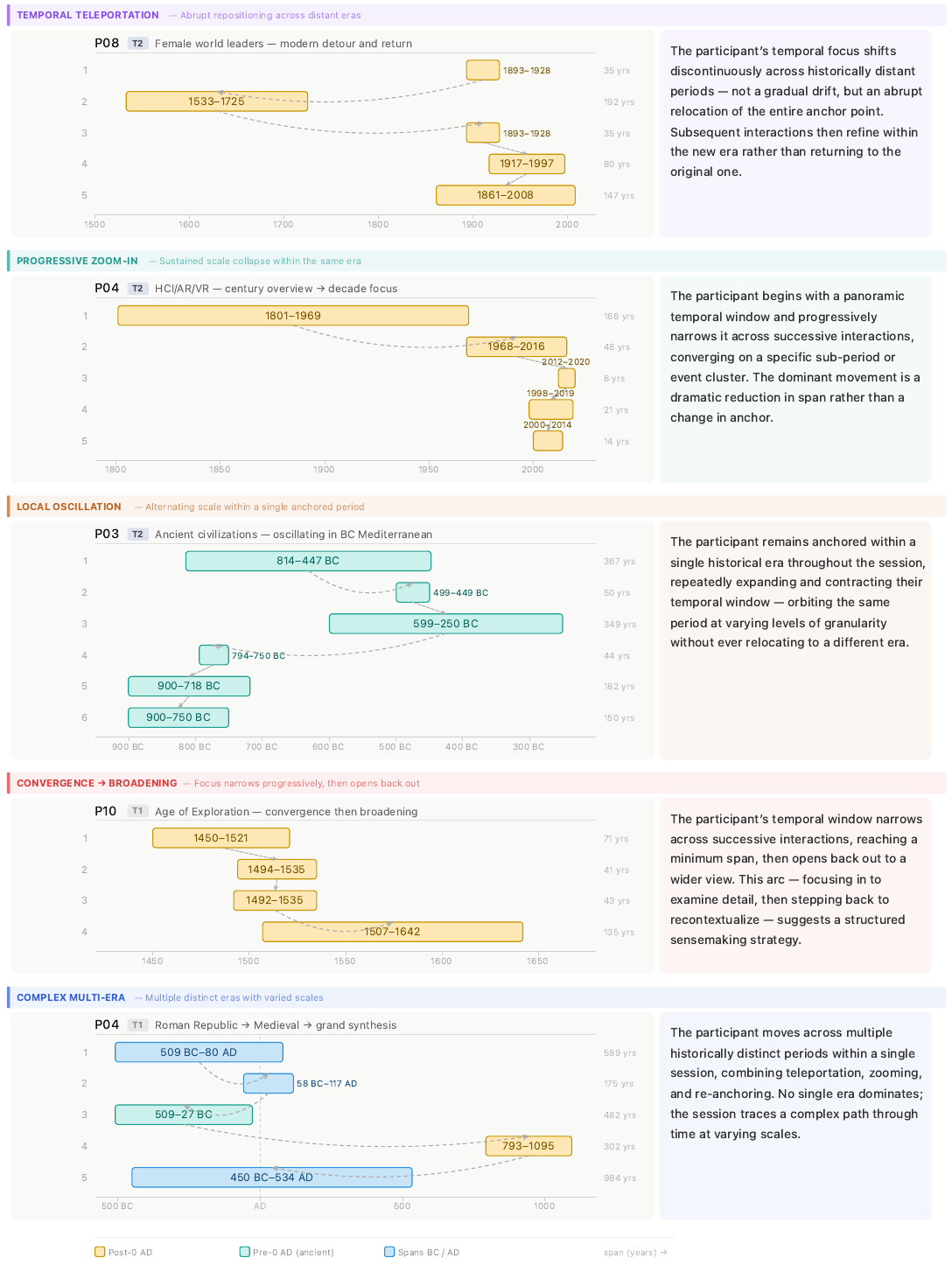}
  \caption{Five recurring temporal navigation patterns identified from user study (see Section~\ref{sec:rq-1}).}
  \Description{Five annotated timeline diagrams illustrating recurring temporal navigation patterns, one per row. Each row shows a participant's sequential temporal range selections as horizontal bars on a year axis, with dashed arrows connecting successive ranges. Bars are color-coded: yellow-orange for post-0 AD, teal for pre-0 AD (ancient), and blue for ranges spanning BC and AD. Row 1 (Temporal Teleportation, P08): five bars jumping discontinuously between 1533–1725 and 1861–2008 on an axis from 1500 to 2000. Row 2 (Progressive Zoom-In, P04): five bars progressively narrowing from 1801–1969 down to 2000–2014 on an axis from 1800 to 2000. Row 3 (Local Oscillation, P03): six teal bars alternating in width between approximately 44 and 367 years, all anchored within 900–250 BC. Row 4 (Convergence then Broadening, P10): four bars narrowing then widening between 1450 and 1650. Row 5 (Complex Multi-Era, P04): five bars spanning from 509 BC to 1095 AD on an axis from 500 BC to 1000 AD, mixing all three color types. Each row includes a text description of the pattern on the right.}
	\label{fig:time-range-shift}
\end{figure*}

\subsection{RQ1: How Do Users Explore Ideas, Events, and Relationships Using KnowledgeTrail?}
\label{sec:rq-1}

Participants used generative timelines to engage in non-linear, branching exploration, frequently shifting between overview and detail, revisiting earlier events, and opportunistically following emerging questions. These exploration patterns differed markedly from the largely linear prompt–response interaction observed when participants used ChatGPT.

To characterize how exploration unfolded across systems, we mapped low-level interaction logs and observable ChatGPT turns to a shared set of system-agnostic exploration intents (e.g., initiating inquiry, expanding an event, connecting events), enabling comparison across interfaces. Figure~\ref{fig:interaction_sequence} visualizes representative interaction sequences for both systems. When using ChatGPT, participants' interactions largely followed a linear pattern, alternating between entering prompts and reading responses, with occasional follow-up questions. In contrast, interactions with KnowledgeTrail were more heterogeneous and interleaved — participants frequently navigated between events, expanded individual nodes for explanation, generated relationships across events, and shifted between levels of abstraction using semantic zoom. Notably, all ChatGPT sessions in which connecting events appeared belonged to KT-first participants,\footnote{Participants assigned to a group where they first used KnowledgeTrail (KT) and then ChatGPT} a pattern we examine further in Section 5.2.3.


Figure~\ref{fig:temporal-distribution} quantifies this contrast. Wilcoxon signed-rank tests show KnowledgeTrail produced significantly more interactions overall (W=2, p=.012, r=.93) and more breadth- and depth-oriented events in raw count (p<.05). Importantly, when controlling for total interaction volume (i.e., comparing the proportion of breadth, depth, and connection interactions within each session), between-system differences were not significant (all p>.07), suggesting that KnowledgeTrail lowered interaction friction rather than reshaping the compositional structure of exploration. 

Figure~\ref{fig:time-range-shift} further reveals that participants varied considerably in the temporal scales at which they explored. Some participants, such as P04, shifted focus across broad historical sweeps spanning centuries, while others, such as P10, began with a narrower temporal window and progressively broadened their scope — a pattern we describe as \emph{\textbf{\hlcb{ConvergencePattern}{convergence-then-broadening}}}. These differences suggest that generative timelines support not only non-linear inquiry but also flexible multi-scale navigation, allowing users to zoom in and out of temporal structure in response to their evolving questions.

Across participants, we identified five recurring navigation patterns: \emph{\textbf{\hlcb{TeleportationPattern}{temporal teleportation}}}, in which participants abruptly repositioned focus across historically distant eras (e.g., P08 shifting from 1893–1928 to 1533–1725 in a single step); \emph{\textbf{\hlcb{ProgressivePattern}{progressive zoom-in}}}, in which the temporal window contracted multiple times within a single era (e.g., P04 collapsing from a general research in HCI spanning 168 years to recent developments in AR/VR research); \emph{\textbf{\hlcb{OscillationPattern}{local oscillation}}}, in which participants repeatedly expanded and contracted their window within a single anchored period without changing their historical anchor (e.g., P03); \emph{\textbf{\hlcb{ConvergencePattern}{convergence-then-broadening}}}, in which focus narrowed progressively before reopening to a wider scope (e.g., P10); and \emph{\textbf{\hlcb{MultiEraPattern}{complex multi-era navigation}}}, in which participants moved across multiple distinct historical periods at varying scales within a single session (e.g., P04). These patterns suggest that generative timelines support not only non-linear inquiry but also flexible multi-scale navigation.

Qualitative examples further illustrate the range of sensemaking that generative timelines supported. Even expert users encountered genuine discoveries: P4, who had deep familiarity with the history of augmented and virtual reality, confirmed that seven of eight generated events "all makes sense" — yet discovered during the session that the first AR system was developed in 1990 by Tom Caudell, a Boeing engineer, and that Caudell had also coined the term `augmented reality.' Generative timelines also supported extended causal narrative-building. Exploring the Roman Empire, P5 noticed a temporal gap between \node{LegendPolitics}{Nero's suicide (68 AD)} and the \node{LegendPolitics}{Construction of the Colosseum (80 AD)} — remarking ``they were able to construct this thing... [although] there was a civil war [after his death]. This is interesting'' — then finding \node{LegendPolitics}{Vespasian becomes Emperor (69 AD)} and realizing: ``So in 68 AD, there was Nero's suicide. And in one year, Vespasian came out and stabilized everything. That's amazing. This emperor must be really good.'' When told time was nearly up, P5 replied: `No, I want to ask further questions about this.'

\subsection{RQ2: How Do KnowledgeTrail Features Shape Exploration and Sensemaking?}

Participants attributed their ability to explore, connect, and make sense of temporally structured information to specific features that supported explanation-on-demand, relational reasoning, and flexible shifts in abstraction.

\subsubsection{Usage Analysis.} 
We analyzed feature usage across 10 participants (excluding P1 and P7 due to recording loss). We focused our analysis on the second session where they explored the topic of their choice, because the second session arguably better reflects their natural usage. Participants used the \events feature in the \textsf{Search Box} most frequently (N=41), while the most used feature in the \textsf{Expand Bar} was \explain (N=24), suggesting participants regularly sought explanations when browsing events. Most participants (10/12) used \relationships (N=11) to explore connections between multiple events.


\subsubsection{Other Features} Participants broadly found the \textsf{Timeline View} and \textsf{Side Panel} useful for exploring events and building comprehensive understanding of topics (9 Strongly Agree, 2 Agree, 1 Neutral), with relationship explanations rated as particularly essential (8 Strongly Agree, 4 Agree). Flexible navigation via \textsf{Side Panel} links helped participants quickly locate relevant content (Fig.~\ref{fig:nav-side-panel}). Semantic zoom was valued for identifying high-level patterns and relationships across the canvas (4 Strongly Agree, 8 Agree).

\subsubsection{Generate Relationship} 

Being able to prompt any two or more events for relational explanation was among the valued features (e.g., P1, P3-5, P10-12) because of its potential to surface connections participants had never considered. During the study, P1, who has a research background in mycology, was pleasantly surprised while using \relationships feature to discover that \node{LegendDiscovery}{Discovery of Vitamin D (1922)} led to the \node{LegendDiscovery}{Discovery of Ergosterol (1928)} (Fig.~\ref{fig:generate-relationship}). 
When no direct relationship existed between selected events, the system acknowledged this and offered contextual framing instead, helping users calibrate their expectations for generated content.
P1 said: 

\begin{displayquote}
``Being able to look at events within events and exploring various events via the relationship feature is great for not only understanding the big picture but also the nitty gritties. \textbf{You also get to understand how two seemingly unrelated things were going side by side}. Only yesterday did I make the connection that Einstein's theory of relativity and Freud's psychoanalytic theory were proposed during the same time. Surely, the two must have read the others' work, but for me, these had always been two separate worlds. So I can see how this system can help a learner gain a more comprehensive understanding of the world and history. If nothing else, \textbf{every user is going to gain a sense of wonder and awe connecting everything together.} Brilliant stuff.''
\end{displayquote}


A post-hoc observation from the interaction analysis further underscores this: no ChatGPT-first participant tried Connecting Events (CE; i.e., asking ``\textit{what are relationships between X and Y}?'') in ChatGPT — their first system — even among those who went on to use CE extensively in KT (e.g., P03, with 4 CE instances in KT). In contrast, all CE interactions in ChatGPT originated from KT-first participants (P05, P06), who translated KT's relational affordance into explicit connective prompts. This suggests that KT exposure primed and elicited relational thinking even as participants moved to a conversational interface — and that relational inquiry does not emerge spontaneously from conversational AI use alone. More broadly, this suggests that temporal-relational affordances may shift users' mental models of what is askable from an AI assistant — a form of interaction priming that extends beyond the immediate session.


\subsection{RQ3: What Potential Uses Do Participants Envision for KnowledgeTrail?}

Within the learning context, P6 suggested enabling learners to depict their current understanding of the history --- by adding events and topics to the timeline --- and have the system verify or correct their understanding of the history.

P5 felt that the way \sys{} allows users to visualize knowledge and connections can be useful for ``presenting your theoretical brain.'' P5 said: ``you have many different ideas, and although those ideas have connections, explicitly making connections has nuances. So it's difficult to explain them, but having those visualizations really helps.''

P4, a graduate researcher whose research is in augmented and virtual reality, noted its potential to support various research activities, including identifying research gap, formulating an explanation of their work's contribution (by highlighting its difference from prior work), and finding relevant theories. P4 noted: 

\begin{displayquote}
``If I'm brainstorming a new research project, it'd be helpful to use this tool. Say, I want to build a new haptic bot, then of course I want to see the history of this research field; what are the available options so far; what are the issues that are raised throughout this process. I would say the tool is already very flexible now, but there can be more advancements to show all the details like papers. For example, if you click an event, it can show all the related papers.''
\end{displayquote}

P10 shared that even though he is not someone interested in history, he ``thoroughly enjoyed the experience of using [KnowledgeTrail]'' and noted that this could be very useful in museums, not only to learn history but also science.

\section{Discussion}

In this work, we introduced KnowledgeTrail, an interactive system that uses LLMs to generate a dynamic timeline for exploratory sensemaking over historical events and their relationships. Our exploratory study suggests that KnowledgeTrail supports curiosity-driven exploration by helping users surface relationships between topics, traverse temporal structure flexibly, and iteratively refine their lines of inquiry. Below we discuss limitations, implications, and directions for future work.

\subsection{Limitations \& Future Work} 

Although participants found the system easy to use, it remains difficult to fully characterize exploratory experience within a single study session. Within brief exploration periods, participants generated a large number of topics, suggesting the need for mechanisms to abstract, collapse, or set aside portions of the timeline no longer in active use. Several participants expressed interest in saving and revisiting workspaces, pointing to the importance of supporting continuity across sessions — for example, allowing users to build upon existing workspaces or selectively merge parallel explorations when connections emerge. Future work should examine how generative timelines are created, revisited, and reconfigured over extended use through longitudinal studies.

Several system-level directions also warrant future attention. The system could be enhanced to maintain context across explored topics, automatically linking newly generated nodes to existing ones to surface connections that users might not prompt directly. The current system also generates 4–8 events per query; greater flexibility in event count and granularity would better accommodate different exploratory goals and topic scopes. Further work is also needed on prompt scaffolding for non-expert users: as Fig.~\ref{fig:thermodynamics-example} illustrates, users sometimes need to reformulate context queries precisely (e.g., changing 'how it evolved' to 'how it evolved after this equation's derivation') to obtain temporally accurate generation—a sensitivity that may be difficult to navigate without guidance.

\subsection{Implications}
\label{sec:discussion-implications}

\indent\textbf{Sensemaking and Exploratory Use.}
Our findings point to broader questions about how generative AI systems could be designed to support open-ended sensemaking rather than the retrieval of predefined knowledge. Participants engaged with KnowledgeTrail as an exploratory medium, using the timeline to follow curiosity, surface relationships, and iteratively refine their understanding of a topic space. This raises important questions about how such systems should scaffold epistemic judgment — particularly when users explore unfamiliar domains, and extends beyond educational contexts to research, journalism, and other forms of knowledge work.

\textbf{Design Implications for Generative Timeline Interfaces.}
The five patterns identified from our user study carry implications for the design of generative timeline interfaces. \emph{\textbf{\hlcb{TeleportationPattern}{Temporal teleportation}}} (abrupt repositioning across distant eras) suggests that interfaces should support lightweight bookmarking or breadcrumb trails so users can recover their prior anchor after large jumps. \emph{\textbf{\hlcb{ProgressivePattern}{Progressive zoom-in}}} (sustained narrowing of the temporal window within a single era) suggests that interfaces should offer progressive scope-narrowing controls rather than relying on free-form pan/zoom, to scaffold sustained focus on sub-periods. \emph{\textbf{\hlcb{OscillationPattern}{Local oscillation}}} (repeated expansion and contraction within an anchored period) suggests that interfaces should make the home era explicit and recoverable, allowing users to orbit a familiar period at varying granularities. \emph{\textbf{\hlcb{ConvergencePattern}{Convergence-then-broadening}}} (focus-then-reflect behavior) argues for explicit `widen scope' affordances (e.g., a contextual prompt or button) after a zoom-in sequence, to support deliberate recontextualization. Finally, \emph{\textbf{\hlcb{MultiEraPattern}{Complex multi-era navigation}}} (traversal across multiple distinct periods at varying scales) suggests that generative timelines should support named region markers or era-level annotations to help users maintain orientation across a session.

\indent\textbf{Mitigating Biases of Temporal Structure.}
Temporal structure is not a neutral sensemaking scaffold. By foregrounding chronological organization, generative timelines may not foreground associative or thematic forms of inquiry that users might otherwise pursue. A related concern is that the visual co-location of events on a shared timeline, combined with AI-generated relationship descriptions, may encourage users to infer causal links that the underlying evidence does not support — temporal proximity on the canvas can be misread as causal proximity. Future systems should consider explicit uncertainty signals that distinguish correlation-based proximity from causally-grounded relationships, and might explore hybrid structures that allow switching between temporal and associative views of the same knowledge space.

\textbf{Designing for Credible Citations.} To examine whether source citations (Fig.~\ref{fig:event-generation} E\&F) were an appropriate mechanism for addressing participants' concerns about factual accuracy, we conducted a brief design probe (N=6; Appendix \ref{appendix:design-probe}) in which participants used a version of KnowledgeTrail with citation cues enabled. Participants generally found citations useful (M=6.3/7) and reported that their presence increased trust in generated content (M=5.0/7), even when links were not actively followed — describing citations as reassuring indicators that the system was not "making things up" rather than as primary verification tools. Participants were more likely to inspect sources when content felt surprising or when the system generated relationships between non-obvious events, suggesting that citations supported selective critical engagement rather than routine checking.

These findings also point to design tensions worth addressing in future systems. Trust in citation cues was fragile: broken links or unfamiliar sources were sufficient to undermine confidence, reflected in moderate ratings for citation quality (M=4.7/7) and relevance (M=5.7/7) despite high perceived utility overall. Participants also differed in what they considered credible — those with academic backgrounds questioned the suitability of Wikipedia, while others found it acceptable — suggesting that configurable source preferences (e.g., academic versus general-web) would make provenance support more adaptable across contexts of use.

\textbf{Extending Serendipitous Exploration Beyond Historical Exploration.}
Generative timelines position timelines not simply as a representations of information, but as a dynamic space for curiosity-driven exploration and sensemaking. One important implication is their potential to support serendipitous discovery, as users encounter unexpected yet meaningful events and connections in the course of exploration. Designing systems that embrace and enhance such serendipity — including how systems surface promising directions without overly steering users — could expand the potential of generative timelines well beyond historical inquiry. In research, for example, a generative timeline might organize papers, hypotheses, experiments, and findings as an evolving knowledge space, allowing researchers to follow particular lines of inquiry while serendipitously encountering connections across otherwise separate literatures or stages of their own research process.

\section{Conclusion}

In this paper, we introduced generative timelines as a new class of interface and instantiated the concept through KnowledgeTrail, demonstrating how it can support open-ended exploration, serendipitous discovery, and relational sensemaking. Through an exploratory study, we examined how users engage with generative temporal representations, surfacing five recurring navigation patterns, a priming effect on relational thinking, and design insights around serendipity and credibility that arise when timelines are no longer fixed artifacts. While prior approaches have ranged from manually authored timelines to NLP- and LLM-based summarization, these efforts treat timelines primarily as static outputs. KnowledgeTrail introduces a different perspective — the timeline as an interactive and generative medium that evolves alongside the user's curiosity — highlighting a path forward for systems that actively scaffold exploration, sensemaking, and discovery.

\bibliographystyle{ACM-Reference-Format}
\bibliography{references}

\appendix

\section{Appendix}
\label{sec:appendix}

\subsection{Implementation Details}

\sys{} is developed as a web application, using ReactFlow, a canvas library, as its main UI. For text generation, we employed OpenAI's GPT-4 model, and for image generation, we used DALL·E 3. All interactions for content generation were executed in real time. While latency may vary due to unstable network connections and differing model selection, the average latency for LLM content generation during the user studies for the interactions were: \textsc{Event Generation}: 8-10 seconds, \textsc{Explanation/Relationship Generation}: 2-4 seconds and \textsc{Image Generation}: 8-10 seconds. There were no complaints about the latency of the \sys{} system during any of the user studies. The prompts utilized for generating content are detailed in Table~\ref{table:llm-prompts} in the Appendix. Source citation functionality was implemented using OpenAI's Web Search feature, which extracts and provides links to source websites (e.g., Britannia, Wikipedia)~\cite{chatgpt-search}.

\newpage

\subsection{User Study}

\begin{table}[h]
\centering
\resizebox{0.48\textwidth}{!}{%
\begin{tabular}{lllll}
\toprule
\textbf{} & \textbf{Gender} & \textbf{Age} & \textbf{AI Tools} & \textbf{Usage Frequency} \\
\midrule
P1 & F & 26 & ChatGPT, Gemini & On a daily basis \\
\midrule
P2 & M & 23 & ChatGPT & On a daily basis \\
\midrule
P3 & M & 27 & ChatGpt, Claude, Gemini advanced & On a daily basis \\
\midrule
P4 & M & 25 & ChatGPT, Github CoPilot, Notion AI & On a daily basis \\
\midrule
P5 & M & 25 & ChatGPT & On a monthly basis \\
\midrule
P6 & F & 25 & ChatGPT, ClovaNote, Otter.ai & On a daily basis \\
\midrule
P7 & F & 22 & ChatGPT & On a weekly basis \\
\midrule
P8 & F & 32 & ChatGPT and Writer & On a weekly basis \\
\midrule
P9 & F & 44 & ChatGPT, Claude and Gemini  & On a daily basis \\
\midrule
P10 & M & 27 & ChatGPT, Claude, Perplexity, Gemini & On a daily basis \\
\midrule
P11 & F & 29 & Chatgpt, Grok AI, Gemini, Monica & On a daily basis \\
\midrule
P12 & M & 26 & ChatGpt, Gemini, Claude, Scite AI, CoPilot & On a daily basis \\
\bottomrule
\end{tabular}
}
\caption{User study participant demographics along with AI tools they have used to learn topics, the frequency of their ChatGPT (Baseline) usage, and to what extent they have used ChatGPT (Baseline) for learning topics.}
\label{tab:participant_info}
\end{table}

\subsection{User Study Results}

\begin{table}[h!]
\centering
\caption{Creativity Support Index (CSI) Results (N=12). The highest value is in \textbf{bold}. The second highest in \underline{underline}.}
\begin{tabular}{lcc}
\hline
Factor & Av.g. Score (SD) & Avg. Factor Count \\
\hline
Enjoyment & \textbf{18.58}/20 (1.51) & 2.00 \\
Exploration & \underline{18.42}/20 (1.90) &  \textbf{3.92} \\
Results Worth Effort & 18.33/20 (1.74) & \underline{3.58} \\
Expressiveness & 17.12/20 (2.20) & 2.92 \\
Immersion & 17.08/20 (2.21) & 1.92 \\
\hline
Overall CSI Score & 85.94/100 (4.32) &  \\
\hline
\label{tab:csi}
\end{tabular}
\end{table}

\begin{figure}[htb!]
    \centering
    \begin{subfigure}[t]{0.47\textwidth}
    \includegraphics[trim=2.3cm 0cm 0cm 0cm, clip=true, width=1\columnwidth]{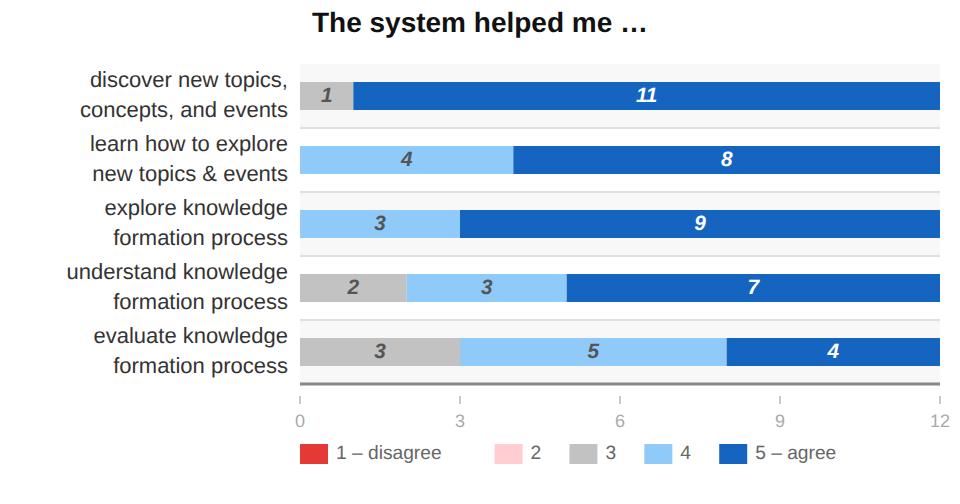}
    \caption{Support for exploring, understanding, and evaluating knowledge formation process (i.e., tracing ideas, events, and their relationships)}
    \label{fig:system-helped-me}
    \end{subfigure}
    \begin{subfigure}[t]{0.47\textwidth}
        \includegraphics[trim=2.3cm 0cm 0cm 0cm, clip=true, width=\textwidth]{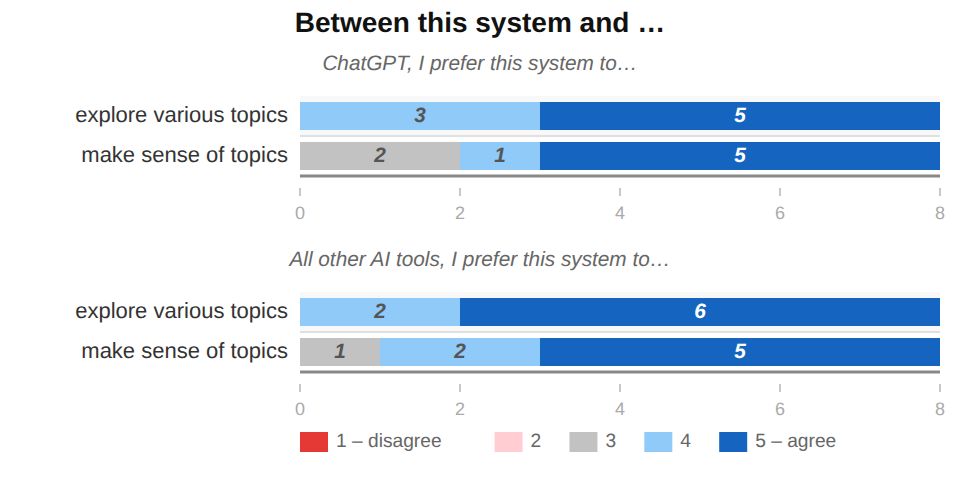}
        \caption{Comparison with ChatGPT and other AI Tools}
        \label{fig:between-this-system}
    \end{subfigure}
    \caption{System Evaluation Results}
    \Description[short description]{This image shows a bar chart based on the survey results. Specifically, it shows how well KnowledgeTrail supports exploring, understanding, and evaluating knowledge formation process and whether participants prefer KnowledgeTrail to ChatGPT.}
    \label{fig:system-eval}
\end{figure}

\subsection{Design Probe: Citation Cues}
\label{appendix:design-probe}

We recruited 6 participants (D1–D6; 2F, 4M) via Upwork, all with prior experience using AI tools for exploratory or informational purposes (e.g., ChatGPT, Perplexity, NotebookLM). Four reported using AI tools daily; two used them a few times per week or month. In a pre-study survey, participants reported moderate trust in AI-generated information (M=4.8/7) and rated citations as highly important for evaluating information (M=6.0/7).

Each session lasted approximately 90 minutes. Participants explored one assigned topic (ROME or EXPLORATION) and one self-selected topic using KnowledgeTrail with citation cues enabled. ChatGPT was not used in this probe. Following exploration, participants completed a post-study survey measuring: perceived usefulness of citation cues (M=6.3/7), influence of citations on trust in generated content (M=5.0/7), perceived citation quality (M=4.7/7), and perceived citation relevance (M=5.7/7). Sessions concluded with a semi-structured interview on when citations felt helpful or unnecessary and suggestions for improving provenance support.

\begin{table*}[h!]
\centering
\caption{Transcript describing the discoveries P5 made while exploring `The History of the Roman Empire' (SA: Study Administrator). It demonstrates progressive discovery and curiosity elicited through exploration.}
\resizebox{0.99\linewidth}{!}{
\begin{tabular}{m{2.8cm} p{6.3cm} p{10cm}}
\hline
Key Moment & Action Description & Transcript \\
\hline
\makecell[{{p{2.8cm}}}]{Noticing \\temporal gap} & \makecell[{{p{6.3cm}}}]{P5 looks at \node{LegendPolitics}{Nero's suicide (68)} and \node{LegendPolitics}{Construction} \\\node{LegendPolitics}{of the Colosseum (80)}. P5 wonders how Rome recovered from instability after Nero's suicide to construct the Colosseum.} & \makecell[{{p{10cm}}}]{P5: ``...The important question is: this (Nero's suicide) happened in 68 and then the construction of Colosseum happened in 80. So they're saying, basically Roman Empire was in a great trouble and then they were able to construct this thing. \textit{This is interesting}.''} \\
\midrule
\makecell[{{p{2.8cm}}}]{Seeking \\causal explanation} & \makecell[{{p{6.3cm}}}]{P5 reads the \textsf{Side Panel} description attributing the Colosseum's construction to the stability and prosperity of Vespasian's rule.} & \makecell[{{p{10cm}}}]{P5: ``Vespasian became the emperor immediately after Nero's suicide. \textit{But, how is it possible?} Because what it is suggesting is, there was Nero's suicide and there was like an instability, there was a civil war, and \textbf{one year after Vespasian became an emperor, that sort of stabilized everything. This emperor must be really good.}''} \\
\midrule
\makecell[{{p{2.8cm}}}]{Recognizing\\ temporal clustering} & \makecell[{{p{6.3cm}}}]{P5 notices a cluster of \node{LegendPolitics}{Nero's suicide (68)}, \node{LegendPolitics}{Year of the Four Emperors (69)}, \node{LegendPolitics}{Vespasian bec-} \\\node{LegendPolitics}{omes Emperor (69)}, and \node{LegendPolitics}{Construction of the C-} \\\node{LegendPolitics}{olosseum (80)}.} & \makecell[{{p{10cm}}}]{P5: ``\textbf{I think that's already a very interesting historical event that we found.} There was great instability. There was like a civil war. This emperor basically normalized everything and made the Roman Empire stable again, which led to the construction of Colosseum in a very short period of time.''} \\
\midrule
\multirow{2}{3cm}{\makecell[{{p{2.8cm}}}]{Curiosity persisting under time pressure}} & \multirow{2}{6cm}{\makecell[{{p{6.3cm}}}]{P5 curious to explore more.}} & \makecell[{{p{10cm}}}]{SA: ``Okay, you have two more minutes. Are you done exploring?''} \\
& & {P5: ``No, I \textbf{want to ask further questions about this}.''} \\
\midrule
\makecell[{{p{2.8cm}}}]{Generating \\follow-up questions} & \makecell[{{p{6.3cm}}}]{P5 looks at \node{LegendPolitics}{Vespasian becomes Emperor (69)} and clicks the \questions button and selects \chip{LegendOther}{What} \chip{LegendOther}{significant actions did he take during his reign?} (Fig.~\ref{fig:explain-and-questions})} & \makecell[{{p{10cm}}}]{P5: ``...\textbf{That's amazing.} The chaos of the Year of Four Emperor. What? Oh, wow. So did that happen within the year? That's amazing.''} \\
\midrule
\makecell[{{p{2.8cm}}}]{Synthesizing \\discovery narrative} & \makecell[{{p{6.3cm}}}]{P5 reads the \textsf{Side Panel} explanation (Fig.~\ref{fig:explain-and-questions}) and summarizes the sequence of events.} & \makecell[{{p{10cm}}}]{P5: ``So that (civil war) happened within a year. That's amazing. So in 68, there was Nero's suicide. And in one year, Vespasian came out and stabilized everything. \textbf{That's amazing. I think this is a very interesting event to report for a presentation}.''} \\
\hline
\label{tab:p5-exploration}
\end{tabular}
}
\end{table*}

\begin{table*}[htb!]
    \caption{Topics in the user study. \textcolor{ACMBrown}{`\{\{ Topic \}\}'} is a placeholder for one of the topics in the Topics column.}
    \label{table:user-study-topic}
    \centering
    \resizebox{0.8\linewidth}{!}{
    \begin{tabular}{p{6.5cm} p{8.5cm}}
    \toprule
 Topics & Task Prompt \\
 \midrule
 & \\
 {\textcolor{ACMBrown}{The History of the Roman Empire: The Roman Empire was one of the most significant and influential civilizations in history, spanning over a thousand years from around 27 BCE to 476 CE (though the Western Roman Empire fell in 476 CE, the Eastern Roman Empire, also known as the Byzantine Empire, continued until 1453 CE.}\newline} & \multirow{5}{7.5cm}{Imagine that you are a student giving a presentation on the following topic in two weeks: \newline

-> \textcolor{ACMBrown}{\{\{ Topic \}\}} \newline

Your goal is to find significant and interesting events that took place throughout \textcolor{ACMBrown}{\{\{ Topic \}\}}. \newline

This is a group project, so, before the presentation, you are meeting your group members to share what you have found from the research, and discuss the story your group should tell in the presentation. You will use this tool to share the findings from your research, so your goal should be:
\begin{itemize}
    \item Identify as many significant topics, events, themes, and questions related to this topic.
    \item Organize the information in a way that would help your group members easily make sense of it.
    \item Brainstorm potential stories you can tell from the discovered topics, events, themes, and questions.
\end{itemize}
} \\\cline{1-1}
 & \\
 {\textcolor{ACMBrown}{The Age of Exploration, also known as the Age of Discovery, was a period of European history that spanned from the 15th to the 17th centuries. It was characterized by extensive maritime exploration, colonization, and the establishment of trade routes that had far-reaching consequences for global history.}} & \\
 & \\
 & \\
 & \\
 & \\
 & \\
 & \\
    \midrule
    \bottomrule
    \end{tabular}
    }
\end{table*}

\begin{table*}[htb!]
    \caption{A list of topics study participants explored as well as a list of topics provided to help select a topic when they were allowed to choose a topic. The \textbf{bold} indicates two pre-determined topics (ROME and EXPLORATION) participants had to explore using either ChatGPT or \sys{}. The \underline{underline} indicates topics four participants (P02, P03, P05, P07, P11) chose from the given list. The \underline{\textit{italic underline}} indicates topics that were not from the given list that the other participants chose.}
    \label{table:topic-of-choice}
    \centering
    \resizebox{0.7\linewidth}{!}{
    \begin{tabular}{p{3cm} p{10.5cm}}
    \toprule
 Domain & Topic \\
 \midrule
History  & \makecell[l]{
- \underline{Ancient Civilizations} \\
- World Wars \\ 
- Renaissance Period \\
- Industrial Revolution \\
- Cold War Era \\
- \textbf{The History of Roman Empire} \\
- Civil Rights Movement \\ 
- Colonialism and Imperialism \\
- Medieval History \\ 
- \textbf{The Age of Exploration} \\
- \underline{The Fall of Empires} \\
- Industrial Revolution \\
- Colonialism and Imperialism \\
- World Wars \\
- History of Science and Medicine \\
- \underline{\textit{Stone Age}} \\
} \\
\midrule
 Science & 
\makecell[l]{
- Theory of Evolution \\
- Discovery of electricity \\
- Medical milestones \\
- Environmental science milestones \\
- Scientific revolution \\
} \\
\midrule
 Physics & 
\makecell[l]{
- Atomic theory \\
- \underline{Thermodynamics} \\
- Cosmology \\
} \\
\midrule
 Computer Science & 
\makecell[l]{
- \underline{\textit{The History of Augmented Reality and Virtual Reality Research}} \\
- \underline{\textit{Accessibility Research in Human-Computer Interaction}}
}\\
\midrule
Philosophy  & 
\makecell[l]{
- Enlightenment philosophy \\
- Postmodernism philosophy \\
- Utilitarianism \\
- Postmodernism \\
- \underline{Metaphysics} \\
- Existentialism \\
- Philosophy of Mind \\
- Eastern Philosophies (e.g., Buddhism, Confucianism) \\
} \\
\midrule
Biology  & 
\makecell[l]{
- Cell biology \\
- Anatomy \\
- \underline{\textit{Mycology}} \\
} \\
\midrule
Other  & 
\makecell[l]{
- \underline{\textit{The Origin of Jazz Music}} \\
- \underline{\textit{Female World Leaders}} \\
- \underline{\textit{Vietnam}} \\
} \\
\bottomrule
\end{tabular}
}
\end{table*}

\begin{table*}[h]
    \caption{Prompts used in \sys{}.}
    \label{table:llm-prompts}
    \centering
    \resizebox{0.85\textwidth}{!}{
    \begin{tabular}[t]{ p{2.1cm}  p{5.2cm} p{4.7cm}  p{5.6cm} }
        \toprule
\textbf{Prompt Type}      
& \textbf{Prompt}   
& \textbf{Example Input(s)} 
& \textbf{Example Response} \\\midrule
\textsc{Events}
& \textsf{Give me \textcolor{ACMOrange}{\{numOfTopics\}} give or take \textcolor{ACMOrange}{\{numOfMargin\}} events that helped to  
        directly support \textcolor{ACMOrange}{\{topic\}} within the broader context of \textcolor{ACMOrange}{\{context\}}. Be sure to focus on \textcolor{ACMOrange}{\{topic\}}.
        The exact year of an event should be given in your response in the format YYYY.
        If the year is BC, use a negative number instead of BC.
        Do not include \textcolor{ACMOrange}{\{topic\}} as one of your events.
        Try to ensure that all events take place generally close together in time (within 500 years).
        The type of event should also be given as only one of the following: [Theory, Discovery, Invention, Politics, Art, Economics, Other]
        IMPORTANT: Sort events by year and ensure that your response in JSON format as described below:  
        \{``events'':[
        \{``Event\_name'':``Event\_name'',
        ``Year'':``Year'',
        ``Type'':``Type''\}
        ]\}
        } 
& \textsf{\textcolor{ACMOrange}{8}, \textcolor{ACMOrange}{2}, \textcolor{ACMOrange}{Age of Discovery}, \textcolor{ACMOrange}{North America} }
& {\small\fontfamily{lmss}\selectfont

\{``events'':[

\{``Event\_name'':``Christopher Columbus' first voyage'',
``Year'':``1492'',
``Type'':``Discovery''\},

\{``Event\_name'':``John Cabot's discovery of Newfoundland'',
``Year'':``1497'',
``Type'':``Discovery''\},

\{``Event\_name'':``Vasco Núñez de Balboa discovers the Pacific Ocean'',
``Year'':``1513'',
``Type'':``Discovery''\},

\{``Event\_name'':``Hernán Cortés conquers the Aztec Empire'',
``Year'':``1521'',
``Type'':``Politics''\},

\{``Event\_name'':``Francisco Pizarro conquers the Inca Empire'',
``Year'':``1533'',
``Type'':``Politics''\},

\{``Event\_name'':``Jacques Cartier's first voyage, discovering Canada'',
``Year'':``1534'',
``Type'':``Discovery''\},

\{``Event\_name'':``Sir Walter Raleigh's expedition to Roanoke'',
``Year'':``1584'',
``Type'':``Discovery''\},

\{``Event\_name'':``Founding of Jamestown'',
``Year'':``1607'',
``Type'':``Politics''\}
]\}
}  \\\hline
\textsc{Explain}        
& \textsf{ \textsc{Explain} \textcolor{ACMOrange}{\{topic\}} within the broader context of \textcolor{ACMOrange}{\{context\}}. Be sure to focus on \textcolor{ACMOrange}{\{topic\}}. The maximum length of this explanation should be 3000 characters. It does not need to be this long. It can be short.} 
& \textsf{\textcolor{ACMOrange}{Age of Discovery}, \textcolor{ACMOrange}{North America}}
& {\small\fontfamily{lmss}\selectfont The Age of Discovery refers to a period in history during the 15th to 17th centuries when European explorers set out to discover new lands and establish trade routes. This era had a profound impact on North America as European powers such as Spain, France, and England began to explore and colonize the continent...}  \\\hline
\textsc{Questions}
& \textsf{I need to learn about \textcolor{ACMOrange}{\{topic\}} in relation to the overall context of \textcolor{ACMOrange}{\{context\}}. Give me a total of 5 questions. Do not add number in front of the questions. Format your response with comma separated values (CSV)}
& \textsf{\textcolor{ACMOrange}{Age of Discovery}, \textcolor{ACMOrange}{North America}} 
& {\small\fontfamily{lmss}\selectfont What were the main motivations behind the Age of Discovery?}, {\small\fontfamily{lmss}\selectfont How did the Age of Discovery impact the indiginous populations of North America?}, {\small\fontfamily{lmss}\selectfont What were some of the major expeditions in North America during the Age of Discovery?}, ... {\small\fontfamily{lmss}\selectfont What were the long-term impacts of the Age of Discovery on North America?}  \\\hline
\textsc{Generate Image}
& \textsf{Create a non-photo realistic, sketch-like image for the concept: \textcolor{ACMOrange}{\{topic\}}. The image should be representative of the following content: \textcolor{ACMOrange}{\{result\}}}
& \textsf{\textcolor{ACMOrange}{Age of Discovery}, \textcolor{ACMOrange}{Description for Age of Discovery}} 
& {\vspace{0pt}\centering\includegraphics[trim=0cm 0cm 0cm 0cm, scale=0.25, clip=true, width=0.25\textwidth]{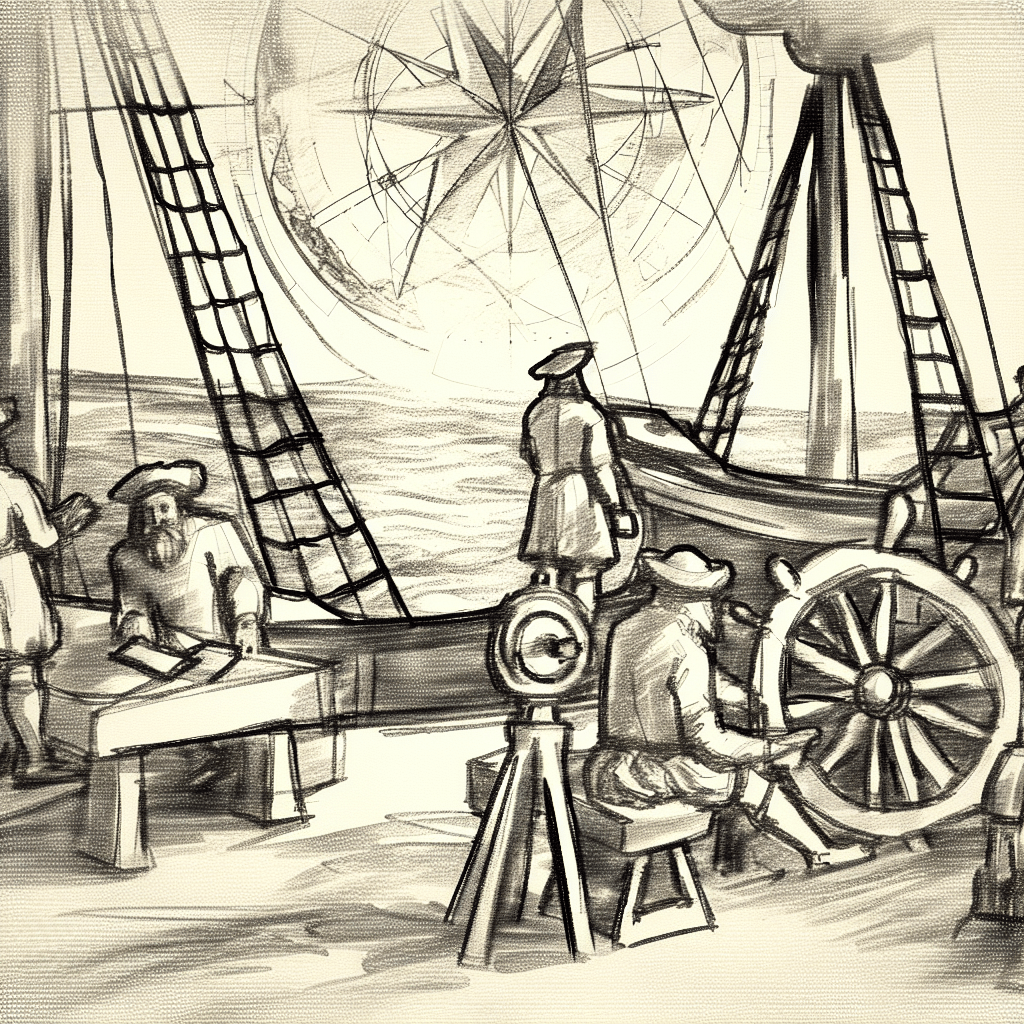}}   \\\hline
\textsc{Generate Relationship}
& \textsf{I need to learn about \textcolor{ACMOrange}{\{topic\}} in relation to the context of \textcolor{ACMOrange}{\{context\}}. In a short paragraph using the following format, tell me how these events relate each other and within the context of \textcolor{ACMOrange}{\{topic\}}: \textcolor{ACMOrange}{\{subevents\}}
    Focus on describing the relationship between each event and \textcolor{ACMOrange}{\{topic\}}. Format: When referencing any event, follow it with a @ and the exact event name as inputted. Surround both the reference and the event name with =}
& \textsf{\textcolor{ACMOrange}{Age of Discovery}, \textcolor{ACMOrange}{North America}, \textcolor{ACMOrange}{[Christopher Columbus' first voyage, John Cabot's discovery of Newfoundland, Vasco Núñez de Balboa discovers the Pacific Ocean, Hernán Cortés conquers the Aztec Empire, Francisco Pizarro conquers the Inca Empire, Jacques Cartier's first voyage, discovering Canada, Sir Walter Raleigh's expedition to Roanoke, Founding of Jamestown]}} 
& {\small\fontfamily{lmss}\selectfont The Age of Discovery was marked by expeditions and conquests that expanded European influence and understanding of the world, particularly North America. Christopher Columbus' first voyage marked the beginning of this era, as he sought a new route to Asia and instead discovered the Americas. Vasco Núñez de Balboa discovering the Pacific Ocean symbolized the exploration of new territories, while Hernán Cortés conquering the Aztec Empire and Francisco Pizarro conquering the Inca Empire represented the conquest and colonization...}  \\\hline
        \bottomrule
    \end{tabular}
}
\end{table*}

\begin{table*}[h]
    \caption{Example events / topics generated given the topic and context}
    \label{table:generated-event-examples}
    \centering
    \resizebox{0.85\textwidth}{!}{
    \begin{tabular}{ p{3cm}  p{3cm} p{10cm} }
        \toprule
\textbf{Topic}
& \textbf{Context}   
& \textbf{Example Events} 
\\\midrule
United States 
& general knowledge
& \makecell[{{p{10cm}}}]{{\node{LegendPolitics}{Signing of the Declaration of Independence (1776)}, \node{LegendPolitics}{Ratification of the U.S. Constitution (1788)}, \node{LegendPolitics}{Louisiana Purchase (1803)}, \node{LegendPolitics}{War of 1812 (1812)}, \node{LegendEconomics}{California Gold Rush (1848)}, \node{LegendPolitics}{American Civil War (1861)}, \node{LegendInvention}{Completion of the Transcontinental Railroad (1869)}, \node{LegendInvention}{Invention of the Phonograph (1877)}}}
\\\hline
United States
& cold war
& {\node{LegendPolitics}{Truman Doctrine (1947)}, \node{LegendPolitics}{Marshall Plan (1948)}, \node{LegendPolitics}{Formation of NATO (1949)}, \node{LegendPolitics}{Korean War (1950)}, \node{LegendPolitics}{Vietnam War (1955)}, \node{LegendPolitics}{Cuban Missile Crisis (1962)}, \node{LegendPolitics}{Detente (1971)}, \node{LegendPolitics}{Reagan Doctrine (1985)}}
\\\hline
United States
& around 19th century
& {\node{LegendInvention}{Invention of the Cotton Gin (1793)}, \node{LegendEconomics}{Louisiana Purchase (1803)}, \node{LegendPolitics}{War of 1812}, \node{LegendPolitics}{Monroe Doctrine (1823)}, \node{LegendEconomics}{California Gold Rush (1848)}, \node{LegendPolitics}{Seneca Falls Convention (1848)}, \node{LegendPolitics}{Civil War (1861)}, \node{LegendInvention}{Completion of Transcontinental Railroad (1869)}}
\\\hline
Theory of relativity
& general knowledge
& {\node{LegendDiscovery}{Discovery of Electromagnetism (1820)}, \node{LegendDiscovery}{Discovery of Speed of Light (1862)}, \node{LegendTheory}{Formulation of Laws of Electromagnetism (1865)}, \node{LegendInvention}{Invention of Quantum Mechanics (1885)}, \node{LegendDiscovery}{Discovery of Radioactivity (1896)}, \node{LegendDiscovery}{Discovery of Electron (1897)}, \node{LegendTheory}{Planck's Quantum Theory (1900)}, \node{LegendDiscovery}{Discovery of Special Relativity (1905)}, \node{LegendDiscovery}{Discovery of General Relativity (1915)}}
\\\hline
        \bottomrule
    \end{tabular}
}
\end{table*}

\begin{table*}[h]
    \caption{Factual Accuracy Analysis}
    \label{table:accuracy-analysis}
    \centering
    \resizebox{0.75\textwidth}{!}{
    \begin{tabular}{ p{3cm}  p{3cm} p{4cm} p{3cm} }
        \toprule
\textbf{Component}
& \textbf{Total \# of responses generated}   
& \textbf{Total \# of accurate responses} 
& \textbf{\% accuracy (averaged over all user studies)} 
\\\midrule
EVENT OCCURRENCE
& 424
& 415
& 98.3\%
\\\hline
EVENT YEAR
& 424
& 416
& 98.1\%
\\\hline
DESCRIPTIONS PANEL
& 48
& 48
& 100\%
\\\hline
RELATIONSHIPS PANEL
& 61
& 52
& 87.0\%
\\\hline
        \bottomrule
    \end{tabular}
}
\end{table*}

\begin{table*}[h]
    \caption{User Study Error Examples. This does not include descriptions generated (DESCRIPTIONS PANEL), as there was no error.}
    \label{table:error-examples}
    \centering
    \resizebox{0.95\textwidth}{!}{
    \begin{tabular}{ p{3cm}  p{2cm} p{5cm} p{7cm} }
        \toprule
\textbf{Component}
& \textbf{User Study}   
& \textbf{Incorrect Generation} 
& \textbf{Explanation of Error} 
\\\midrule
EVENT OCCURRENCE
& P06 
& Disability Theory: From Margin to Mainstream (2002)
& This Book/Article/Paper does not actually exist
\\\hline
EVENT YEAR
& P04 
& Launch of PlayStation VR (2020)
& Actual year is 2016
\\\hline
RELATIONSHIPS PANEL
& P03
& The First Peloponnesian war and the peace of Callias further solidified Athenian power, making the construction of the Parthenon a testament to Athens' strength and cultural acheivements. 
& 
This statement is correct for the peace of Callias. The Peloponnesian war (431BC), however, is actually more recent than all of these events and was a war between Athens and Sparta for the Greek world. Sparta won this war so it did not help solidify Athenian power or lead to the construction of the Parthenon.
\\\hline
        \bottomrule
    \end{tabular}
}
\end{table*}

\begin{figure*}
    \centering
    \includegraphics[trim=0cm 0cm 0cm 0cm, clip=true, width=\textwidth]{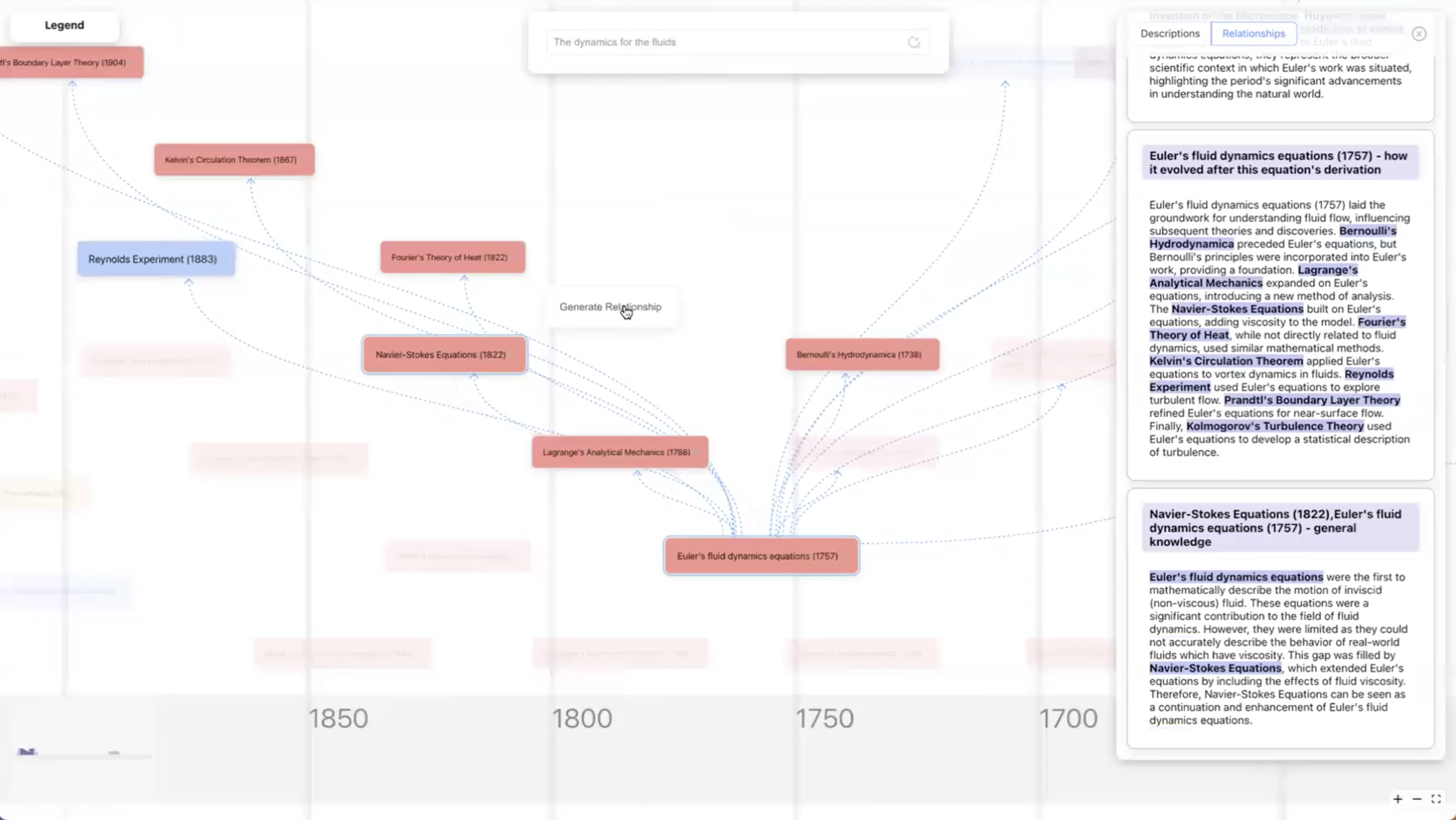}
    \caption{A screenshot of P5's exploration of the historical development of \hlcb{events}{\textsf{thermodynamics}}. As shown, P5 explored events prior and after \hlcb{events}{\textsf{Euler's fluid dynamics equation}}. To create the events that \hlcb{events}{\textsf{Euler's fluid dynamics equation}} led to, P5 had to change the context from \hlcb{button}{\texttt{how it evolved}} to \hlcb{button}{\texttt{how it evolved after this equation's derivation}}, showing the need for some prompt engineering in order to get \sys{} to behave the way users want.}
    \Description{A screenshot of the KnowledgeTrail interface showing P5's exploration of the historical development of thermodynamics. The timeline canvas displays multiple interconnected event nodes spanning approximately 1700 to 1900, including Euler's fluid dynamics equations (1757), Navier-Stokes Equations (1822), Reynolds Experiment (1883), and several others connected by dashed arrows. The Side Panel on the right shows two open results: one titled "Euler's fluid dynamics equations (1757) — how it evolved after this equation's derivation" and another comparing Navier-Stokes Equations and Euler's equations, with key event names bolded and hyperlinked. The Legend Panel in the upper left lists event types including Theories, Discoveries, and Inventions.}
    \label{fig:thermodynamics-example}
\end{figure*}


\begin{table*}[t]
\small
\caption{Intent coding scheme applied to interaction log analysis (KnowledgeTrail; \textcolor{ktblue}{\textbf{KT}}) and turn-level analysis (ChatGPT; \textcolor{gptbrown}{\textbf{GPT}}). The same intent categories were applied to both systems. Categories with no ChatGPT equivalent were coded for KnowledgeTrail sessions only.}
\label{tab:coding-scheme}
\resizebox{0.9\linewidth}{!}{
\begin{tabular}{P{1.6cm} P{2.6cm} l P{9.6cm}}
\toprule
\textbf{Dimension} & \textbf{Intent Category} & \textbf{System} & \textbf{Operationalization} \\
\midrule

\multirow{8}{=}{\textbf{Breadth}}
& \multirow{2}{=}{Initiating Inquiry}
  & \textcolor{ktblue}{\textbf{KT}}
  & Entry of a topic into the \textsf{Search Box} followed by clicking the \events button — either with a general context (e.g., ``\textsf{General Knowledge}'') or a specific contextual constraint (e.g., ``\textsf{World War II}''). Coded at the first interaction on a new topic thread. \\
\cmidrule(l){3-4}
& & \textcolor{gptbrown}{\textbf{GPT}}
  & Open-ended or constrained topic query entered as a new prompt (e.g., ``\textsf{Give me the most significant events for X}'', ``\textsf{Tell me about the history of X}'' or ``\textsf{Tell me about X focusing on Y}''). Coded when the user establishes a new topic, with or without a built-in constraint. \\
\cmidrule(l){2-4}

& \multirow{2}{=}{Broadening Exploration}
  & \textcolor{ktblue}{\textbf{KT}}
  & Clicking the \events button from the \textsf{Expand Bar} on an already-established event node to generate multiple related events and expand the scope of the current inquiry. \\
\cmidrule(l){3-4}
& & \textcolor{gptbrown}{\textbf{GPT}}
  & A manually composed query asking for related or adjacent events within an already-established topic (e.g., ``\textsf{What other events happened around this time}?''). \\
\cmidrule(l){2-4}

& \multirow{2}{=}{Generative Questions (generate)}
  & \textcolor{ktblue}{\textbf{KT}}
  & Clicking the \questions button from the \textsf{Search Box} or \textsf{Expand Bar} to prompt the system to produce a set of five candidate follow-up questions. Coded at the moment the button is clicked, before any question is selected. \\
\cmidrule(l){3-4}
& & \textcolor{gptbrown}{\textbf{GPT}}
  & Explicitly asking the system what questions one could or should ask about a topic (e.g., ``\textsf{What questions should I be asking about X?}''). \\
\cmidrule(l){2-4}

& \multirow{2}{=}{Overview (Zoom Out)}
  & \textcolor{ktblue}{\textbf{KT}}
  & Zooming out past the semantic zoom threshold (0.4) to collapse event nodes into color-coded dots, or clicking an event type in the \textsf{Legend Panel} to filter and overview matching nodes across the canvas. \\
\cmidrule(l){3-4}
& & \textcolor{gptbrown}{\textbf{GPT}}
  & \textit{No equivalent.} ChatGPT has no spatial canvas or zoom affordance. \\

\midrule

\multirow{6}{=}{\textbf{Depth}}
& \multirow{2}{=}{Deepening Understanding}
  & \textcolor{ktblue}{\textbf{KT}}
  & Clicking the \explain button from the \textsf{Expand Bar} on a specific event node to generate a detailed explanation of that event in the \textsf{Side Panel}. Focus is on a single event already present on the timeline. \\
\cmidrule(l){3-4}
& & \textcolor{gptbrown}{\textbf{GPT}}
  & Follow-up query requesting more detail on a specific event already mentioned in the conversation (e.g., ``\textsf{Tell me more about Y}''). The user must retype or reference the event name explicitly. \\
\cmidrule(l){2-4}

& \multirow{2}{=}{Generative Questions (select)}
  & \textcolor{ktblue}{\textbf{KT}}
  & Selecting and executing one of the five generated questions to pursue a more focused line of inquiry. Equivalent to clicking the \explain button. Always preceded by a Generative Questions (generate) interaction in the same session. \\
\cmidrule(l){3-4}
& & \textcolor{gptbrown}{\textbf{GPT}}
  & A descriptive follow-up query on a specific aspect of a topic, analogous to clicking \explain in the \textsf{Search Box} or \textsf{Expand Bar} (e.g., ``\textsf{Can you describe what X was about}?'' or ``\textsf{What happened during Y}?''). \\
\cmidrule(l){2-4}

& \multirow{2}{=}{Detail (Zoom In)}
  & \textcolor{ktblue}{\textbf{KT}}
  & Zooming in past the semantic zoom threshold to expand collapsed event nodes back into full cards, narrowing the visible temporal scope to examine a specific period or cluster in greater detail. \\
\cmidrule(l){3-4}
& & \textcolor{gptbrown}{\textbf{GPT}}
  & \textit{No equivalent.} ChatGPT has no spatial canvas or zoom affordance. \\

\midrule

\multirow{4}{=}{\textbf{Connection}}
& \multirow{2}{=}{Connecting Events}
  & \textcolor{ktblue}{\textbf{KT}}
  & Selecting two or more event nodes on the canvas and clicking the \relationships button to request a relational explanation between them, displayed in the \textsf{Side Panel} under the \tab{TabColor}{\textsf{Relationships}} tab. \\
\cmidrule(l){3-4}
& & \textcolor{gptbrown}{\textbf{GPT}}
  & Explicit relational query naming two or more specific events and asking how they are connected (e.g., ``\textsf{How are X and Y related?}''). The user must independently identify and name both events. \\
\cmidrule(l){2-4}

& \multirow{2}{=}{Revisiting Prior Context}
  & \textcolor{ktblue}{\textbf{KT}}
  & Navigating back to a previously explored event using \textsf{Side Panel} links, clicking a highlighted event name in the \tab{TabColor}{\textsf{Relationships}} tab to re-center the view on it. \\
\cmidrule(l){3-4}
& & \textcolor{gptbrown}{\textbf{GPT}}
  & Scrolling up in chat history to re-read a prior response, or submitting a new query that explicitly re-references a previously discussed event or topic. \\

\bottomrule
\end{tabular}
}
\end{table*}

\end{document}